\documentclass[useAMS,usenatbib]{mn2e}
\usepackage{amssymb}
\usepackage{graphicx}
\DeclareGraphicsExtensions{.jpg,.pdf,.png,.eps,.pdf}
\usepackage{url}
\usepackage{color}
\usepackage{tabulary}
\usepackage{multirow}
\usepackage{amsmath}
\usepackage{amsfonts}
\usepackage{ulem}
\graphicspath{ {pictures/} }
\newcommand{\beq}{\begin{equation}}
\newcommand{\eeq}{\end{equation}}
\newcommand{\simlt}{\mathrel{\hbox{\rlap{\hbox{\lower4pt\hbox{$\sim$}}}\hbox{$<$}}}}
\newcommand{\simgt}{\mathrel{\hbox{\rlap{\hbox{\lower4pt\hbox{$\sim$}}}\hbox{$>$}}}}

\newcommand{\erg}{\;\mathrm{erg}}
\newcommand{\s}{\;\mathrm{s}}
\newcommand{\dd}{\partial}
\newcommand{\Msol}{\;\mathrm{M}_{\odot}}
\newcommand{\cm}{\;\mathrm{cm}}
\newcommand{\AU}{\;\mathrm{AU}}
\newcommand{\pc}{\;\mathrm{pc}}
\newcommand{\yr}{\;\mathrm{yr}}
\newcommand{\Myr}{\;\mathrm{Myr}}
\newcommand{\keV}{\;\mathrm{keV}}

\def\apjl{ApJL}
\def\apj{ApJ}
\def\mnras{M.N.R.A.S.}
\def\aap{A\&A}
\def\nat{Nat.}
\def\araa{Ann. Rev. A\&A}

\def\apjs{ApJ Supp.}
\def\aj{AJ}

\def\aapr{A\&A Rev.}

\def\prd{prd}

\title[The first HMXBs]{Formation, disruption and energy output of Population III X-ray binaries}

\author[T. Ryu et al.]{Taeho Ryu$^{1}$\thanks{email:
    taeho.ryu@stonybrook.edu}, Takamitsu L. Tanaka$^{1,2}$, Rosalba Perna$^{1,3}$\\ 
  $^1$ Department of Physics and Astronomy, Stony Brook University,
Stony Brook, NY 11794-3800, USA\\
$^{2}$Department of Physics, New York University, 4 Washington Place, New York, NY 10003, USA\\
$^{3}$ Adjunt Fellow of JILA, 440 UCB, Boulder, CO 80309-0440, USA }
\begin{document}

\maketitle

\label{firstpage}

\begin{abstract}
The first astrophysical objects shaped the cosmic environment
by reionizing and heating the intergalactic medium (IGM).
In particular, X-rays are very efficient at heating the IGM
before it became completely ionized, an effect that can be measured through the redshifted
21 cm line of neutral hydrogen.
High-mass X-ray binaries (HMXBs), known to be prolific X-ray sources
in star-forming galaxies at lower redshifts, are prime candidates for driving
the thermal evolution of the IGM at redshifts $z\simgt 20$.
Despite their importance, the formation efficiency of HMXBs from the first
stellar populations is not well understood---as such, their collective X-ray emission
and the subsequent imprint on the 21 cm signature are usually evaluated using free parameters.
Using $N$-body simulations, we estimate the rate of HMXB formation
via mutual gravitational interactions of nascent, small groups of the first stars (Population III stars).
We run two sets of calculations: one in which stars form in small groups of five
in nearly Keplerian initial orbits, and another in which two such groups collide
(an expected outcome of mergers of host protogalaxies).
We find that HMXBs form at a rate of one per $\simgt 10^{4}\Msol$ in newly born stars,
and that they emit with a power of $\sim 10^{41} \erg \s^{-1}$ in the $2-10$ keV band 
per solar mass per year of star formation. 
This value is a factor $\sim 10^{2}$
larger than what is observed in star forming galaxies at lower redshifts;
the X-ray production from early HMXBs would have been even more copious,
if they also formed \textit{in situ} or via migration in protostellar disks.
Combining our results with earlier studies suggests that early HMXBs
were highly effective at heating the IGM and leaving a strong 21 cm signature.
We discuss broader implications of our results,
such as the rate of long gamma-ray bursts from Population~III stars
and the direct collapse channel for massive black hole formation.
\end{abstract}


\section{Introduction}
\label{sec:intro}

A major outstanding goal in cosmology is to piece together the history
of the Universe between Cosmic Dawn, the emergence of the first stars and galaxies,
and the end of reionization, when the radiation from these objects had ionized the intergalactic medium (IGM).
Advances in numerical techniques,
combined with exquisite measurements of the ``initial'' conditions (at a redshift $z\approx 1000$; \citealt{Hinshaw+13}, \citealt{Planck15}),
have led to remarkable simulations \citep[e.g.][]{Abel+02, Turk+09, Stacy+10, greif11, BrommYoshida11} of the conditions leading
up to the former milestone, occurring at $z\simgt 30$, when the Universe was $\approx 100\Myr$ old.
However, reconstructing the subsequent several hundred Myr of cosmic history has proved far more challenging,
due to the difficulties in reliably modeling the numerous forms of feedback from
the first astrophysical objects \citep[e.g.][]{Springel+05, Stinson+06, Sijacki+07}.

In particular, X-rays from the first galaxies can act as a powerful source of feedback \citep[e.g.][]{Venkatesan+01, Machacek+03}
that exerts influence over a wide range of distance scales.
Because hard X-rays (energies $\simgt 1\keV$) have mean free paths
comparable to the Hubble horizon, they can isotropically heat and partially reionize the early IGM
(e.g. \citealt{Peng2001,Venkatesan+03, Ricotti2004, Pritchard2007}).
In fact, they are expected to be the dominant agent in heating the IGM.
Such heating may suppress star formation \citep{Ripamonti+08} and massive black hole (BH) growth \citep{Tanaka2012}
inside low-mass dark matter haloes by raising the Jeans and filtering masses of the IGM \citep{Gnedin00, NaozBarkana07}.
On galactic and circum-galactic scales, soft X-rays ($\sim 0.1-1\keV$)  can affect the formation of stars
and possibly massive black holes (BHs) by promoting the formation of molecular hydrogen via electron-catalyzed reactions
\citep[e.g.][]{Haiman+96,Kuhlen2005, Latif+15, Inayoshi2015}.

In addition to their suspected roles in early galaxy evolution,
X-rays are important also because they can leave an observable signature
that can be exploited to probe the cosmological epoch in question \citep{PritchardJ2008}.
Their thermal impact on the early Universe should be measurable
through the  redshifted $21\cm$ transition line of neutral hydrogen,
which is observed in emission or absorption depending on the relative
temperature of the IGM with respect to that of the cosmic microwave background (CMB).
Several studies have investigated how forthcoming observations of the sky-average amplitude
and power spectrum of the relic $21\cm$ line \citep[e.g.][]{Bowman+08, DARE12, LOFAR13, SCIHI14}
could be used to constrain the astrophysical agent (or agents) responsible for heating the early IGM.

There are sound reasons to expect that the first galaxies produced X-rays in abundance,
and rapidly heated the IGM.
There are two dominant X-ray sources in present-day
galaxies---both powered by gas accretion onto BHs,
and both plausibly prominent shortly after Cosmic Dawn:
gas feeding massive BHs shining as active galactic nuclei (AGN),
and X-ray binaries, powered by a stellar-mass BH gradually cannibalizing a companion star.
Estimates of the mass accumulated by nuclear BHs prior to $z\sim 6$ \citep{Shankar+09, Salvaterra+12},
the existence of very massive BHs at $z\simgt 6$
\citep[e.g.][]{Fan+01, McGreer+06, Willott+07, Willott+09, Mortlock+11, Venemans+13, Banados+14}, 
as well as the observed \citep[e.g.][]{Shen+07} and
theoretically expected \citep{Shankar+09, Tanaka14} increase in their duty cycles toward higher redshifts,
all hint that X-ray AGN may have been much more common during this epoch.
Likewise, high-mass X-ray binaries (HMXBs) dominate
the X-ray emission of star-forming galaxies;
the low metallicity and rapid baryonic mass accretion of the earliest galaxies
both lend credence to the notion that they were rife with HMXBs.
Theoretical models suggest that either type of X-ray source 
could heat the IGM to above the CMB temperature as early as $z\sim 30$,
and that this transition should be measurable by the planned $21\cm$ experiments.

At present, there are too many theoretical uncertainties to determine from the future data
which type of X-ray source---AGN or HMXBs---was
responsible for driving the thermal evolution of the $z\simlt 30$ IGM.
Modeling the early AGN X-ray emission is particularly difficult,
because the conditions for triggering AGN activity are not fully understood even at low redshifts \citep[e.g.][]{HopkinsQuataert10, Treister+12};
because of the uncertainty in the fraction of X-ray photons that is
released into the IGM as opposed to being trapped inside the accretion flow
or reprocessed into the infrared \citep[e.g.][]{Madau+14, Pacucci+15};
and because the epoch, initial masses, and birthplaces
of the massive BH ``seeds'' are not yet constrained by observations \citep[e.g.][]{Volonteri10, Haiman13, TanakaLi14}.
Similarly, studies usually estimate the X-ray contribution from early HMXBs
by simply inferring empirical relations between X-ray luminosity and star formation rate (SFR)
in local galaxies (and modeling the SFR using a semi-analytic cosmological model),
or combining such relations with one or more free parameters
\citep[e.g.][]{xraysource2, Tanaka+15}.

The goal of this study is to alleviate the uncertainties in the formation rate and X-ray output
of HMXBs in the early Universe, by using $N$-body simulations of
nascent groups of the first (Population III, henceforth Pop~III) stars.
We choose the properties of the star groups in our simulations to reflect 
those found in hydrodynamical simulations of Pop~III star formation at $z\sim 20$ (\citealt{Greif12}, \citealt{stacy13}).
We follow the formation and dynamical evolution of compact binaries
over thousands to millions of years,
including the effects of the background gravitational potential and dynamical friction.
This allows us to compute the fraction of Pop~III
stars that form stable, compact binaries, and eventually undergo an X-ray bright phase.
The end result is an estimate of the formation rate of HMXBs in the first
protogalaxies, as well as the amount of X-rays they generate per unit star formation.
To our knowledge, this is the first published estimate of this type.

Our simulations predict a binary formation rate which is similar to what is observationally
inferred in present-day galaxies.
However, we derive a HMXB energy output (normalized to the star formation rate)
that is a factor $\sim 10 - 150$ higher than in present-day star-forming galaxies,
if the HMXB duty cycle is similar to the one in the local Universe.
We find that the X-ray output does not change significantly within the wide variety
of simulation setups considered---such as different orientations for
collisions between star groups, and ambient gas density---and submit that this is a robust estimate.

The findings of this study can be used as model inputs in estimating
the $21\cm$ global signature and power spectrum, but have wider applications.
As stated above, the X-ray output of the first galaxies are also of interest
for studying feedback on smaller scales, such as subsequent star formation
and massive BH formation.

Our work is also relevant for predicting the rates of
long-duration gamma-ray bursts (LGRBs) from Pop~III stars.
LGRBs are important probes that can shed light
on the Universe out to $z>10$ \citep[e.g.][]{Toma2011}.
According to the collapsar model \citep{MacFadyen1999},
progenitors of LGRBs require rapid rotation of the He core
and removal of the H envelope.
Both criteria are satisfied by Pop~III HMXBs,
and it is plausible that massive Pop~III stars in binary systems
are dominant LGRB progenitors in the early Universe \citep{Bromm2006}.
Our results on HMXB formation rates can therefore be used
to predict and interpret observations of high-redshift LGRBs.

The paper is organized as follows. We start in \S\ref{sec:dynamics}
by discussing the problem to be solved---beginning with the equations
of motion, followed by the description of our $N$-body code,
our choices for the initial conditions,
and how the data is interpreted for HMXB formation.
We present our results in \S\ref{sec:results}.
In \S\ref{sec:discussion}, we discuss the implication of our
work for the X-ray output of the first galaxies, as well
as for other topics such as LGRBs and SMBH formation.
We conclude with a summary of our findings in \S\ref{sec:summary}.

\section{Stellar Dynamics}\label{sec:dynamics}

Here, we provide an overview of our simulations---namely:
the equations of motion that are solved to
simulate the dynamical evolution of the star groups;
the numerical scheme we use to solve the equations;
the different types of initial conditions we adopted,
as well as the reasoning behind our choices;
and finally, how the results are interpreted for HMXB formation.

\subsection{The equations of motion}
\label{subsec:theequationofmotion}

Our $N$-body code computes the motion of $N$ objects with (generally different) masses
$m_{i}$, moving under their mutual gravitational influence,
a dissipative dynamical friction force,
and a  background gravitational potential.   
We numerically integrate the equations of motion 
\begin{equation}\label{eq:eom}
\frac{d^{2}}{dt^{2}}\vec{r}_{i}=
\vec{a}_{{\rm g}, i}+\vec{a}_{{\rm df}, i}+\vec{a}_{{\rm bg}, i}.
\end{equation}

The first term on the right-hand side of equation (\ref{eq:eom}) is the specific force due to
Newtonian gravity,
\begin{equation}
\vec{a}_{{\rm g}, i}
= -\sum_{j\neq i}
G~m_{j}~
\frac {\dd ~S(r_{ij})}{\dd ~r_{ij}}~
\frac{\vec{r}_{i}-\vec{r}_{j}}{r_{ij}},
\end{equation}
where $G$ is the gravitational constant, 
$\vec{r}_{i}$ is the displacement of the $i_{{\rm th}}$ star
from the center of the host dark matter halo,
and $r_{ij}\equiv |\vec{r}_{i}-\vec{r}_{j}|$.

We adopt the Plummer softening kernel $S(r_{ij})$ \citep[e.g.][]{Galacticdynamics},

\begin{equation}\label{eq23}
S(r_{ij})=-\frac{1}{\sqrt{r_{ij}^{2}+\epsilon^{2}}}\,,
\end{equation}
where we take $\epsilon=R_{\odot}$.

The second term on the right-hand side of equation (\ref{eq:eom}), $\vec{a}_{{\rm df}, i}$,
is the specific drag force due to dynamical friction.
For collisionless systems, the standard Chandrasekhar formula for
dynamical friction is \citep{Galacticdynamics},
\begin{equation}\label{eq:df1}
\vec{a}_{i}=-4\pi ~\ln\Lambda ~f(X_{i}) ~ \frac{G^{2} m_{i}}{v_{i}^{3}}~\rho(r_{i})~\vec{v}_{i},
\end{equation}
where 
\begin{equation}\label{eq:df2}
f(X_{i})\equiv {\rm erf}(X_{i})-\frac{2}{\sqrt{\pi}} ~X_{i} ~\exp\left(-X_{i}^{2}\right),
\end{equation}

$v_{i}$ is the speed of the $i_{th}$ star with respect to the background,
$X_{i}\equiv v_{i}/(\sqrt{2} \sigma_{v})$,
$\sigma$ is the velocity dispersion,
$\ln\Lambda$ is the Coulomb logarithm and
$\rho(\vec{r}_{i})$ is the local gas density.

We adopt the modified formula for gaseous medium used in \cite{dyformula1}.
This prescription incorporates behaviors found in numerical simulations for
subsonic and supersonic regimes \citep{dyformula2, dyformula3}.
The specific drag force vector always points opposite to the direction
of motion, and is given by:
\begin{equation}\label{eq24}
a_{{\rm df},i}^{\rm (gas)}=-4\pi ~G^{2}~m_i~\rho(\vec{r}_i)~\frac{1}{v_i^{2}}\times f^{\rm (gas)}({\mathcal M}_i), 
\end{equation}
with 
\begin{equation}\label{eq25}
f^{\rm (gas)}({\mathcal M}_i)=\begin{cases}
	0.5~\ln\Lambda \Big[{\rm erf}\left (\frac{{\mathcal M}_i}{\sqrt{2}}\right)
	-\sqrt{\frac{2}{\pi}}{\mathcal M}_i~ \exp\left(-\frac{{\mathcal M}_i^{2}}{2}\right)\Big]
	\\ \hspace{100pt} 0\leqslant {\mathcal M}_i \leqslant 0.8;\\
	1.5~\ln\Lambda \Big[{\rm erf}\left (\frac{{\mathcal M}_i}{\sqrt{2}}\right)
	-\sqrt{\frac{2}{\pi}}{\mathcal M}_i~ \exp\left(-\frac{{\mathcal M}_i^{2}}{2}\right)\Big]
	\\ \hspace{93pt} 0.8\leqslant {\mathcal M} \leqslant \mathcal{M_{\rm eq }};\\
	\frac{1}{2}\ln\Big(1-\frac{1}{{\mathcal M}_i}\Big)+\ln\Lambda
	\\ \hspace{115pt}  {\mathcal M}_i > \mathcal{M_{\rm eq}}.\\
\end{cases}
\end{equation}
Above, ${\mathcal M}_i\equiv v_{i}/c_{\rm s}$ is the Mach number,
and $c_{\rm s}$ is the sound speed.
We use $\ln\Lambda=3.1$ and the corresponding value of
$\mathcal{M_{{\rm eq}}}\approx1.5$.

In our simulations, the motion of the stars with respect to the background gas
is supersonic.
In this regime, the characteristic dynamical friction timescale,
for a circular Keplerian orbit of two bodies with for $m_1 \gg m_2$ and
$v_2\gg c_{\rm s}$, is

\begin{equation}\label{eq11}
\tau_{\rm df}\sim\frac{E_{\rm orb}}{P_{\rm df}}
\sim\frac{1}{80} \sqrt{\frac{m_{1}^3}{m_{2}^{2}G}}\frac{1}{\rho(r_{2}) r_{12}^{3/2}}\,,
\end{equation}
where $P_{\rm df}$ is the frictional dissipation power ($m_{2}a_{{\rm
    df},2}v_{2}$), and $E_{\rm orb}$ the orbital energy
($Gm_{1}m_{2}/2r_{12}$).

The third and last term, $\vec{a}_{{\rm bg}, i}$,  is the specific force due to
the background potential, which is dominated by gas. 
The background potential provides an
additional inward force whose functional form depends on the density profile.
For simplicity, here we use a constant density and explore different values in our simulations.
The force due to the background potential is then
\begin{equation}
\label{eq17}
\vec{a}_{{\rm bg}, i}=-4 \pi G\rho \vec{r}_{i},
\end{equation}
where here $\vec{r}_{i}$ is the vector pointing from the 
center of the halo to the $i$-th star.

The equation of motion is solved iteratively, with
the positions, velocities and accelerations
of each star updated at every time step.
We describe our computational method below.

\subsection{Code Description}
\label{subsec:codedescription}

We perform 3-dimensional, $N$-body simulations with 4th-order \& 5-stage
Runge-Kutta-Fehlberg methods (RKF45 method, \citealt{Fehlberg}) using
adaptive time steps. The RKF45 is a very precise and stable
integration method among the large class of Runge-Kutta schemes,
particularly by adapting the Butcher tableau for Fehlberg's 4(5)
method.

We solve equation (\ref{eq:eom}), as described in the preceding text,
updating the position and velocity components of the stars while
treating the background density of the gas as a uniform and static distribution.

To ensure numerical precision, our computational scheme varies the
value of each subsequent time step analytically, so that
numerical errors for each variable in the simulation do
  not exceed $10^{-13}$ times the size of the variable. 
In some cases, however, this method leads to excessive computational effort for
calculating relatively trivial interactions.
For example, near the pericenter of hyperbolic or highly elliptical encounters, 
the time steps become increasingly small in a runaway fashion
to compensate for the steep rise in acceleration and associated errors. 
In order to avoid such situations,
we implement the following two numerical shortcuts to keep computation times tractable.

The first shortcut is to use analytic approximations for very close 2-body encounters.
This is justified in cases where
pairs of stars are sufficiently close to each other
and isolated from the other stars in the simulation,
so that
(i) the gravitational pull from the other stars and the background potential
are negligible compared to the mutual gravitational pull of the pair,
and
(ii) the orbital motion is supersonic and the dynamical friction force 
can be treated as a linear perturbation to the 2-body Keplerian problem.
The code employs analytic approximations for any close stellar pairs
that satisfy these conditions, and reverts to the RKF algorithm when the conditions
stop being satisfied.

We derived the following approximations for the semi-major axis and the eccentricity:
\begin{equation}\label{eq15}
a(t)=\left[
a(t_{0})^{-3/2}+ \beta~\rho~\sqrt{\frac{G}{\widetilde{\mu} w(q)}}~(t-t_{0})\right]^{-2/3},
\end{equation}
\begin{equation}\label{eq16}
e(t)=\sqrt{1-\frac{a(t)}{a(t_{0})}(1-e(t_{0})^{2})}\,.
\end{equation}

Above, $\widetilde{\mu}$ is the reduced mass, $a(t_{0})$ and $e(t_0)$ are
respectively the semi-major axis and the eccentricity at $t=t_{0}$,
$\beta$ is a dimensionless constant, and $w(q)$ is a function of the
mass ratio that is symmetric about $q=1$. Note that the variable $a$
with subscript indicates the specific force, while without subscript
it represents the semi-major axis. To derive the
  analytical expression for the time-evolution of the orbital distance
  in equation (\ref{eq15}), we integrated the equation of the motion
  of a star under the influence of a dynamical friction torque.
  Equation (\ref{eq16}) then follows from the definition of orbital
  eccentricity in terms of orbital energy and angular momentum.
Note that as the radial distance decreases, the eccentricity increases.
Here,
$\beta$ and $w(q)$ are free parameters and we use $\beta=0.035$ and
$w(q)=q^{1.4}+(1/q)^{1.4}$. The stellar coordinates and velocities are
recovered as functions of $a$ and $e$ using standard expressions for
Keplerian orbits \citep{Galacticdynamics}.

Because we are dealing with systems with more than three bodies
of different masses,
stars frequently form hierarchical triple systems whose motions
are affected by the Kozai mechanism \citep{kozai}.
Since our analytic solutions above do not account
for such changes in mutual inclination, 
we limit the use of these solutions to situations where  $\tau_{\rm outer}<\tau_{\rm Kozai}$,
where $\tau_{\rm outer}$ is the dynamical timescale for the outer pair in the hierarchical triple,
and $\tau_{\rm Kozai}$ is the time scale for the Kozai mechanism,
\begin{equation}
\label{eq19}
\tau_{\rm Kozai}\sim \frac{m_{1}}{m_{2}}\frac{P_{1,2}^2}{P_{1,3}}(1-e_{1,2}^{2})^{1.5}.
\end{equation}
The subscript 1 above indicates the primary star of the inner compact
binary along with the satellite star denoted by the subscript 3, while
the hierarchical tertiary star is indexed by the subscript 2,
and $P$ is the orbital period.

Our second shortcut for keeping the simulation runtimes manageable
is to set a minimum value for the time step.
We choose a physically motivated value,  $10^{-6}\times \tau_{\rm dyn, \,min}$,
where $\tau_{\rm dyn, \,min}$ is the smallest value of the dynamical time between
any two stars in the simulation (that are not being treated by the analytic shortcut above)
at a given time step.
This procedure is necessary when there are three or more stars interacting
at small separations, in which case the analytic approximations above cannot be used.
We note that such situations are rare compared to the places where the analytic shortcut
is applicable.

\subsection{Determining HMXB formation}
\label{subsubsec:DeterminingHMXB}
It is assumed that a HMXB has formed if 
both of the following criteria are satisfied:
\begin{enumerate}
\item \textit{One of two stars forming a binary turns into
  a black hole.} In order to determine which stars turn into
black holes, we need to compare the typical lifetime
($\tau_{{\rm life}}$) of a massive star with the time ($t_{{\rm run}}$)
in the simulation (taken to coincide with the time at which stars are
born). If $\tau_{{\rm life}}>t_{\rm run}$, the star is marked as a
black hole in the simulation. For main-sequence stars, it is possible
to estimate this lifetime from the mass-luminosity relation, that is
$E_{{\rm star}}\sim m$ and $L_{{\rm {\rm star}}}\sim m^{p}$, leading to
$\tau_{{\rm life}}\sim m^{1-p}$ with $2<p<3$. However, due to the
uncertain value of $p$ for Pop III stars, we rather prefer to use here
the nuclear time scale of Pop III stars estimated by \citet{Schaerer}
and \citet{Marigo}. They calculate the H-burning nuclear time scale for these
stars with a stellar evolution code.
We assume that a BH forms if the stellar mass is greater than $8\Msol$.
\item \textit{The two stars are close enough so that the accretion occurs
  through Roche-lobe overflow (RLOF)}.
This criterion is simply written as $R_{{\rm star}}\geq
R_{{\rm RL}}$, where $R_{{\rm star}}$ is the stellar radius, and 
$R_{{\rm RL}}$ is the Roche-lobe radius of the most massive star calculated from the
center of the star to the inner Lagrange point. An approximate
analytic formula to the Roche-lobe radius of star 1 for a wide range
of the mass ratio is \citep{Rocheradius},
\begin{equation}\label{eq7}
\frac{R_{{\rm RL},1}}{r}=\frac{0.49q^{2/3}}{0.6q^{2/3}+\ln(1+q^{1/3})}\,,
\end{equation}
where $r$ is the orbital distance and $q=\frac{m_{2}}{m_{1}}$ is the
mass ratio. 
Although the above expression was derived in a study of binaries
in circular orbits, 
\citet{masstransfer} found that the Roche lobe radius does not differ
very much for eccentric binaries. We therefore apply equation (\ref{eq7})
to all binaries found in our simulation.

\end{enumerate}

If a binary satisfies these two criteria, the binary is marked as a
HMXB. Since the two criteria are independently checked
at every time step,
which criterion is satisfied first is not important.

\subsection{Setup and Initial Conditions}
\label{sec:initialsetup}

We design our simulations with the $20\simlt z\simlt 30$ Universe in mind.
This is the redshift range where the dark matter haloes reach
virial temperatures $\sim 1000-2000\;{\rm K}$, the expected condition for Pop~III formation,
at the highest rates.

The lifetimes of the stars are several Myr, which is comparable
to the timescales on which the host haloes undergo mergers
with other Pop~III forming haloes \citep{Tanaka14}.

The initial positions of the stars are generated quasi-randomly via a Monte Carlo
realization as follows,
motivated by the assumption that they formed inside a shared Keplerian gas disk
at the center of the host dark matter halo.
Their initial radial positions in the disk plane are chosen randomly from a uniform distribution whose
amplitude depends on the characteristic size scale of the star-forming cloud (see below).
Their azimuthal distribution is chosen to be nearly uniformly distributed,
so that the azimuthal angular position of the $n$-th star is
given by $360^{\circ}(n/N)\pm5^{\circ}$, where $N$ is the total number of stars in the group.
This choice is made to minimize the radial gravitational pull between the stars,
so that they do not immediately fly apart.
Naturally, the azimuthal positions do not remain evenly spaced, but become mixed quickly.
Finally, we allow the stars to be displaced out of the plane of the disk.
Their positions perpendicular to the plane are chosen randomly, so that their
vertical displacement out of the disk plane is no more than 
$5\%$ above or below their initial radial displacement from their shared center of mass.
The initial velocities of the stars are assigned to be circular, parallel to the disk plane, and Keplerian 
at the instant the simulation begins.

For this study, we investigated two different size scales of star forming regions.
\begin{enumerate}

\item \textit{Large scale}: Following the results from \citet{stacy13},
where Pop~III protostars are formed
inside star-forming regions with a size of a few thousand AU,
we have performed simulations where the stars are placed
inside a region of size $\sim 2000\AU$.

We explore two different values of uniform, constant gas density,
$10^{6}~{\rm cm}^{-3}$ (denoted $n_{6}$ hereafter) and $10^{4}~{\rm cm}^{-3}$
($n_{4}$). 

For gas of primordial composition, the molecular weight
$\mu$ can vary between $0.6$ and $1.2$, depending on the
ionization fraction. We adopt $\mu=1$ for simplicity;
this choice does not qualitatively affect our results,
since our values for $n$ are selected arbitrarily
with the goal of exploring the qualitative dependence on $n$.

Each simulation is run for $5\Myr$.
The masses of the stars follow the initial mass function (IMF)
they provide with $\alpha=0.17$ ($\frac{dN}{dM}=M^{-\alpha}$),
$M_{\text{max}}=140\Msol$ and
$M_{\text{min}} =0.1\Msol$. For these simulations, we consider
$N=5$ stars.

\item\textit{Small scale}: In the simulations by
\citet{Greif12}, multiple protostars formed several AU apart from each other.
We therefore run a second set of simulations,
where stars form within a 10~AU radius.
Since \citet{Greif12} do not provide a slope for the IMF,
we use $\alpha=0.17$ as above.
That study found that more than half of the mass accreted during the protostar
phase goes to the most massive protostar in the group.

To mimic this behavior, we first generate a star with $M_{\text{max}}=200\Msol$,
then generate each subsequent star with 
$M_{\text{max}}=200\Msol - $ [the sum of the masses of the previously generated stars].

Just as with the large-scale case, we run simulations with number densities  $n_{6}$ and $n_{4}$.
Because this case is the more relevant one for forming HMXBs,
we explore several different configurations:
cases with a star group of $N=5$ stars,
a star group of $N=10$ stars,
and collisions between two groups of $N=5$ stars.
The last case is motivated by the fact that Pop~III-forming minihaloes
undergo frequent mergers, which suggests that the nascent star groups
themselves undergo close encounters.

\end{enumerate}

\section{Results}

Here, we summarize the findings of our $N$-body simulations, 
focusing in particular on the properties of the most compact binaries
found for each set of runs.

\label{sec:results}
\subsection{Large scale}

We ran 20 simulations for $n_{6}$ and 23 for $n_{4}$.

For $n_{6}$, in 12 out of 20 runs, the two most massive stars
(${\rm S}_{1}$ and ${\rm S}_{2}$, where ${\rm S}_{i}$ is the $i$-th most massive
star) form the most compact binary---we denote such a binary with the notation ${\rm B}_{12}$.
Similarly, binaries of type ${\rm B}_{13}$
(i.e. made up of the most massive star ${\rm S}_{1}$ and
the third most massive star ${\rm S}_{3}$)
form the most compact binary in 6 of the runs.

\begin{figure*}
	\centering
	{\includegraphics[width=8.5cm]{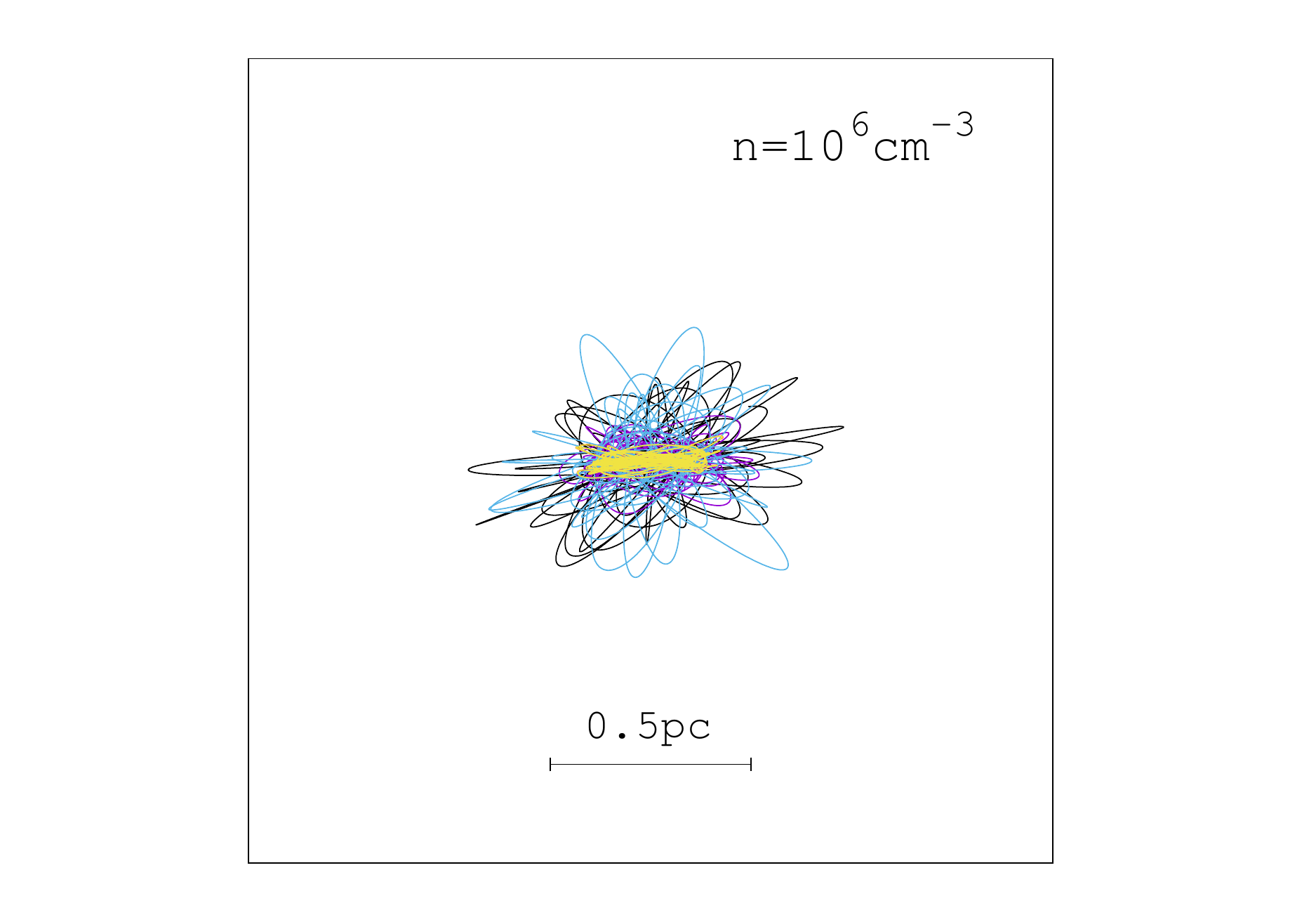}}
	{\includegraphics[width=8.5cm]{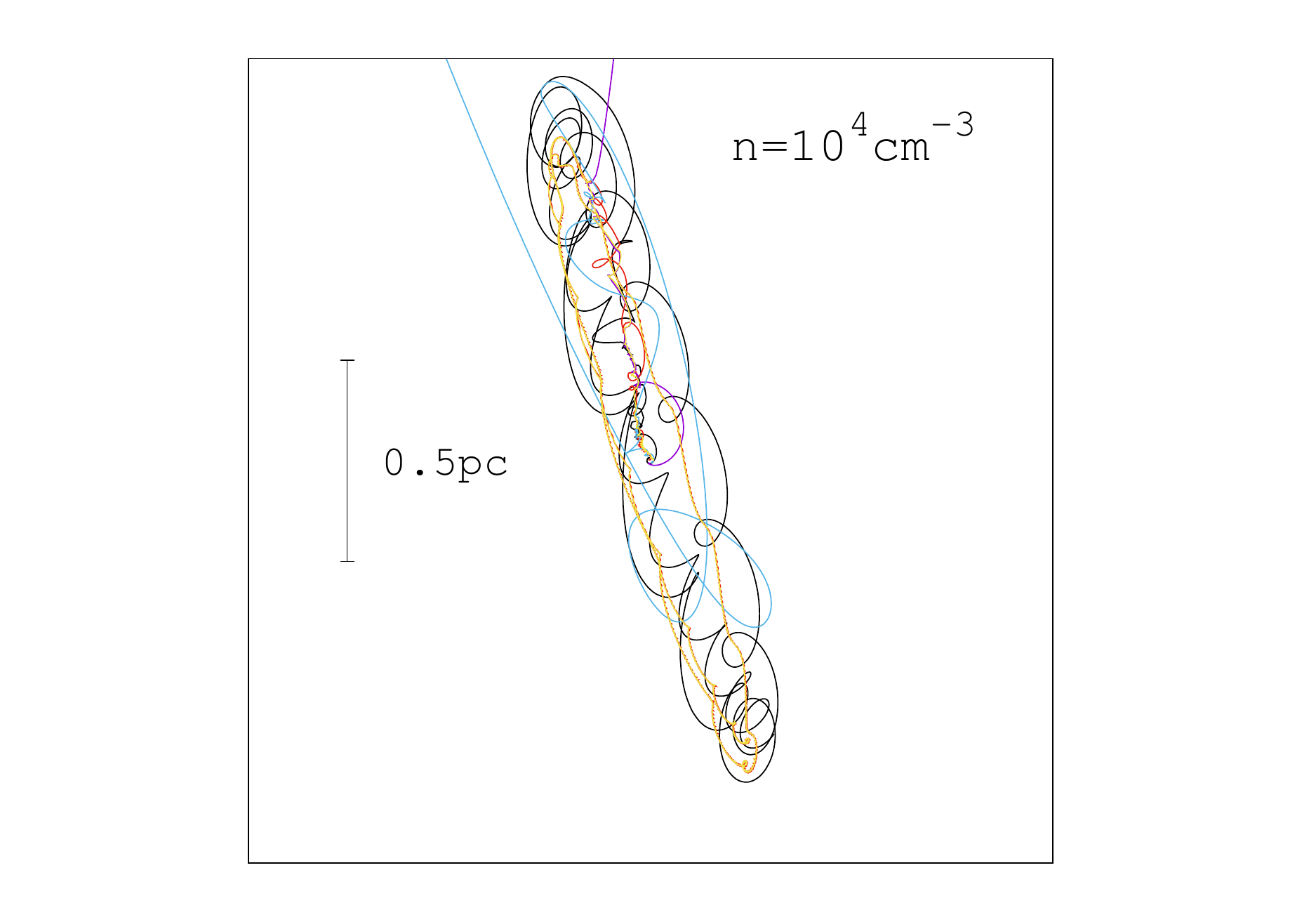}}
	\caption{{\it Left}: Sample trajectories for large scale
          calculations with the high number density $n=10^6\cm^{-3}$ (case $n_{6}$).
           As shown in the figure, stars tend to stay near the center of the halo and
           their overall motions are oblate-spheroidal in shape.
           {\it Right}: 
          Sample trajectories for the low number density $n=10^4\cm^{-3}$ ($n_{4}$). Even
          though stars including binary systems remain within a certain
          distance range, they are not as close as the stars in the
          higher density calculation,
          leading to less frequent three-body interactions.
          Furthermore, one can notice that
          a few stars (blue line and purple line) are kicked off. }
	\label{fig:1}
\end{figure*}

The left panel of Figure~\ref{fig:1} shows a sample set of stellar trajectories
for one of the large scale $n_6$ runs.
Throughout the simulation, the compact binary tends to 
remain near the center of the halo, as less massive
stars repeatedly undergo 3-body interactions with the binary.
The fact that they do not stray far from the center of the halo is a combined effect of the
background potential and the gravitational potential of the massive binary.
It is these 3-body encounters that cause the most
compact binary in the simulations to end up as type ${\rm B}_{12}$ or ${\rm B}_{13}$.
In two of the runs, the most compact binaries after $5\Myr$ consist of less massive stars.

The average value of the semimajor axis after $t=5\Myr$ in the large
scale, $n_6$ runs is $\langle a_{t=5\Myr}\rangle=270\AU$, and the
minimum value of $a$ across the 20 runs is 60~AU.  These values for
$a$ are much larger than that necessary for RLOF to take place, $\sim
0.07$~AU. Their characteristic dynamical friction time scales are
roughly $10^{13}$~yrs, which is much longer than their lifetimes (see
equation \ref{eq7} when $R_{{\rm RL}}=R_{{\rm star}}$ and equation
\ref{eq11}).
We therefore conclude that in systems of $N=5$ stars
and $n\sim 10^{6}~{\rm cm}^{-3}$, with the stars initially separated at hundreds of AU,
HMXBs are unlikely to form.

\begin{table}
	\centering
          \setlength\extrarowheight{5pt}
	\begin{tabulary}{1\linewidth}{c c c}
		\hline
		Large scale & $n_{6}$ & $n_{4}$\\
		\hline
		number of runs &			20					&	23\\
		$P({\rm B}_{12})~[P({\rm B}_{12}+{\rm B}_{13})]$ &			0.60(0.90)					&	0.56(0.74)\\
		$\langle a_{t=5{\rm Myr}}\rangle$[AU]   &  	270			&  	340				  \\
		\hline
	\end{tabulary}
		\caption{Summary of the large scale calculations for
          $n=10^{6}~{\rm cm}^{-3}$ ($n_{6}$) and $n=10^{4}~{\rm cm}^{-3}$
          ($n_{4}$). $P({\rm B}_{12})$ denotes the fraction of the simulations in which the most
          compact binary consists of the two most massive stars.
          We also list $P({\rm B}_{12}+{\rm B}_{13})$, the fraction where the
          most compact binary consists of the most massive star paired with either the second or third most massive star.
	 Also shown is the average semi-major axis of the most compact binary at $t = 5\Myr$.
          }
	\label{tab:tab1}
\end{table}

\begin{figure}
	\centering
	\makebox[0cm]{\includegraphics[width=8cm]{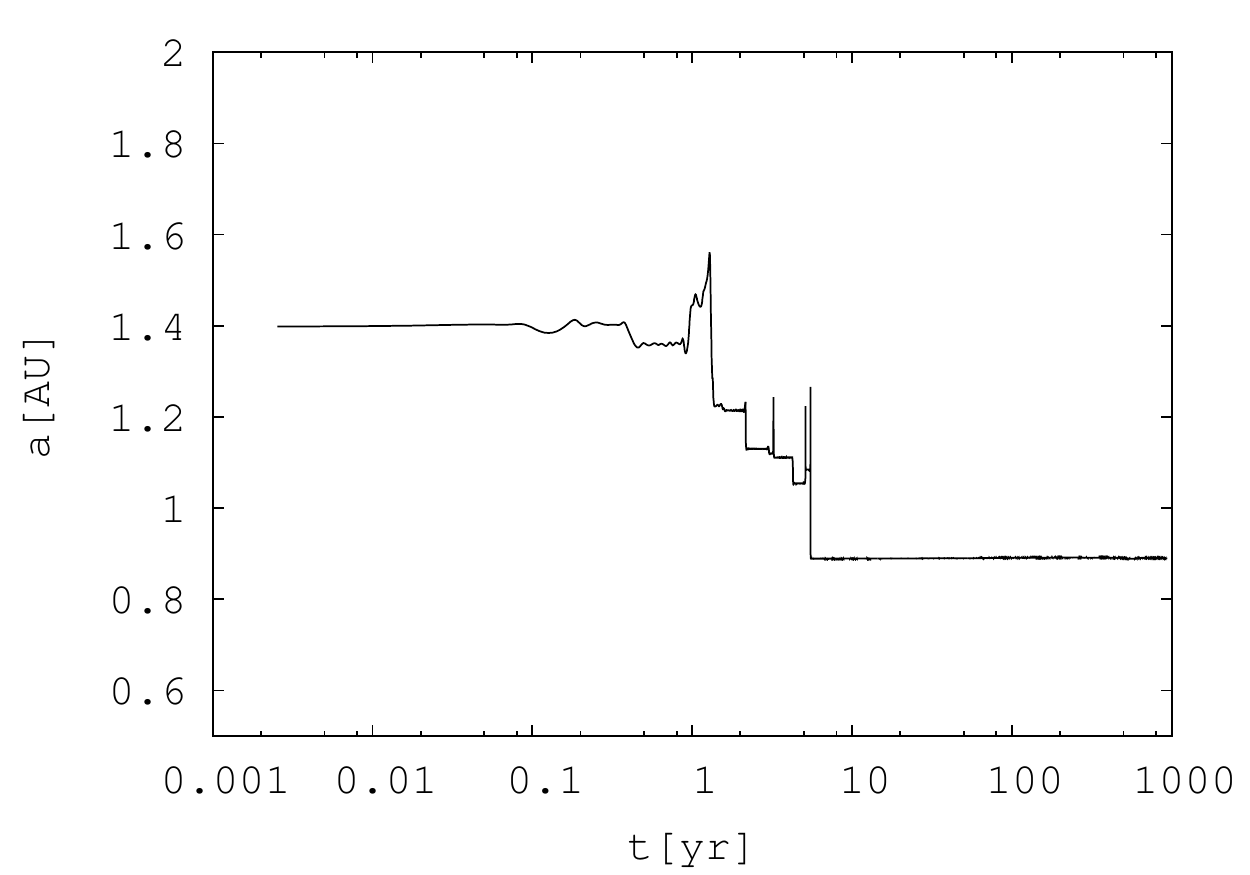}}
	\caption{
	Evolution of the semi-major axis of a typical binary in a system.
           The sharp variations
          are due to stellar scatterings, which mostly result in a
          hardening of the binary. After the multiple system gets
          stabilized and isolated (which happens after $\approx 10$ years), the decrease rate of the semi-major
          axis depends on dynamical friction alone.}
	\label{fig:7}
\end{figure}

For the $n_{4}$ case, the fraction of runs where the most compact binary
after $5\Myr$ is type ${\rm B}_{12}$ or ${\rm B}_{13}$ are similar to the $n_6$ simulations:
$13/23$ for ${\rm B}_{12}$, and $17/23$ for ${\rm B}_{12}$ \textit{or} ${\rm B}_{13}$.

However, the dynamical evolution is quite different, as can be seen by the right panel
of Figure~\ref{fig:1}. The lower gas densities lead to weaker forces due to dynamical friction and background potential,
and stars (especially less massive ones) tend to be scattered farther from their initial position,
as far as $\simgt 1\pc$. This also results in less frequent three-body interactions in general, and
may explain the larger average value of the semi-major axis after $5\Myr$,
$\langle a_{t=5{\rm Myr}}\rangle=340\AU$.
The formation of HMXBs appears even more intractable for the $n_4$ case.

Sample stellar trajectories for the $n_6$ and $n_4$ large scale cases are shown in
Figure~\ref{fig:1}, and a summary of the results is given in Table \ref{tab:tab1}.

\subsection{Small scale}
\label{subsec:smallscale}

\begin{table}
	\centering
                \setlength\extrarowheight{4pt}
	\begin{tabulary}{0.7\linewidth}{c c c}
		\hline
		Small scale & $n_{6}$ & $n_{4}$\\
		\hline
		runs &			86					&	86\\
		$\langle a_{t=500{\rm yr}}\rangle$ [AU]   &  	1.37					&  	1.42						  \\
		$\tau_{{\rm df}}$[yr]	 &     $\sim10^{13}$		 &   $\sim 10^{15}$			  \\
		Companion stars 		&	\multicolumn{2}{c}{3rd massive star, $11\sim12 \Msol$}\\
		$P_{{\rm HMXBc}}$ &    0.070	 &  0.070			  \\
		$F_{\rm HMXB}$ [$10^{-4}\Msol^{-1}$] &    4.6	 &  4.6		\\
		\hline
	\end{tabulary}
			\caption{Summary of results for simulations of 5-body
			groups forming on small scales.
                  $\langle a_{t=5000~{\rm yr}}\rangle$
                  indicates the semi-major axis at $t=5000$ yr, while
                  $\tau_{{\rm df}}$ represents the dynamical friction
                  timescale required for $a_{t=5000~{\rm yr}}$ to
                  shrink to $a_{{\rm RL}}$ (see equation \ref{eq11}).
                  $P_{\rm HMXBc}$ is the fraction of runs in which a  HMXBc forms,
                  and $F_{\rm HMXB}$ is the number of HMXBc formed across
                  all simulations, normalized by the total mass of the stars in the simulations.                  
                  }
	\label{tab:tab2}
\end{table}

\begin{figure*}
	\centering
	{\includegraphics[width=8.7cm]{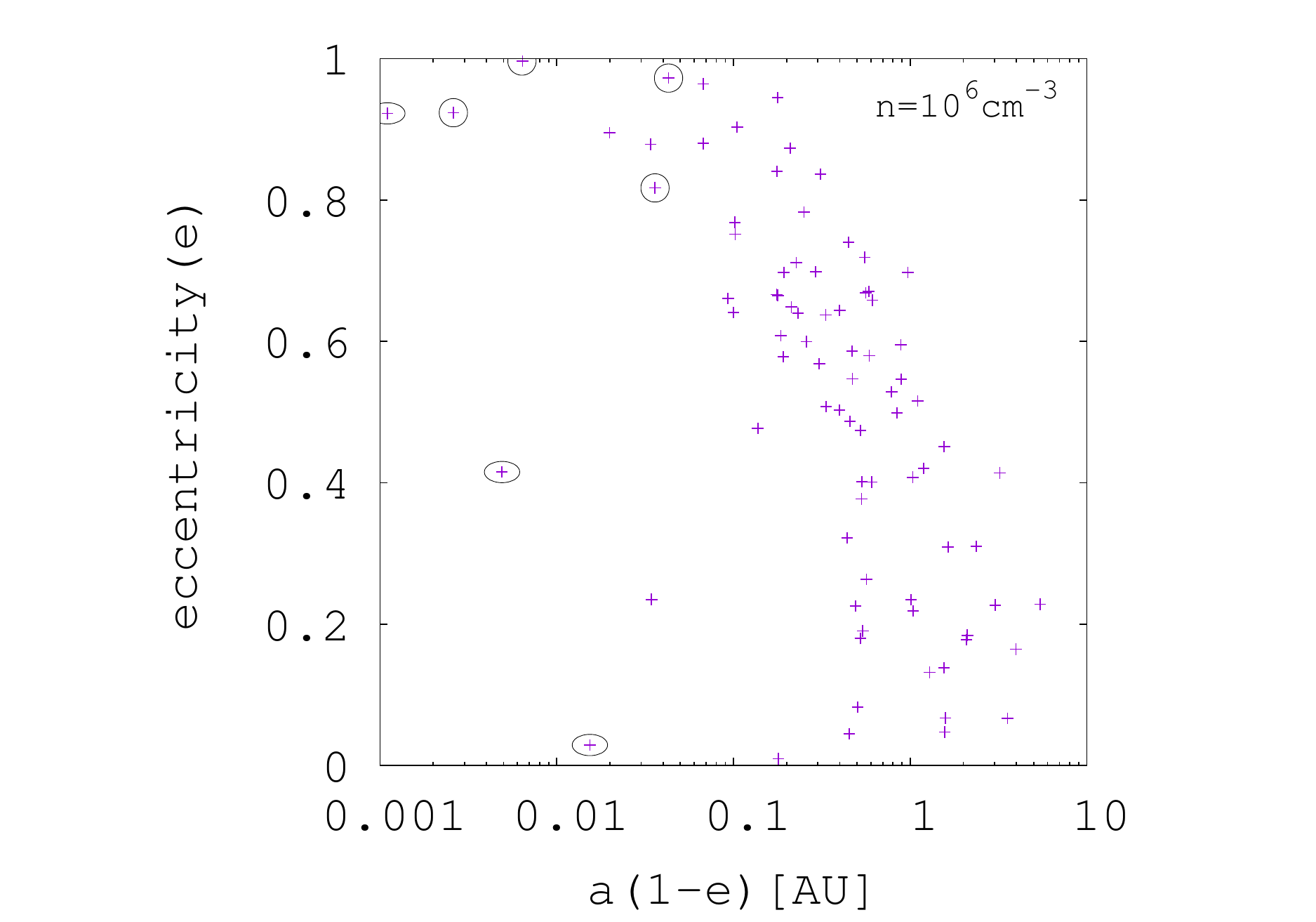}}
	{\includegraphics[width=8.7cm]{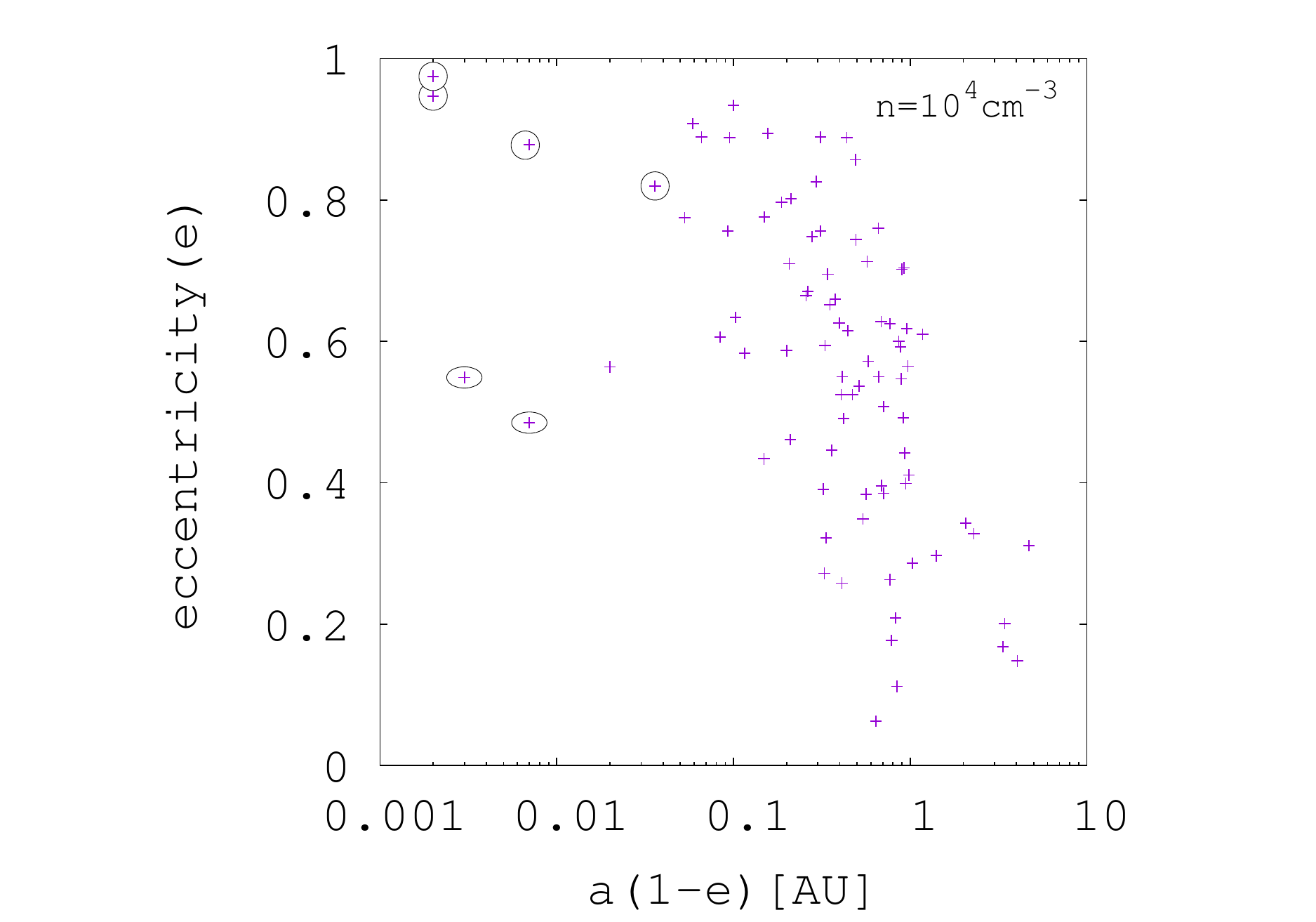}}
	\caption{ {\it Left}: Pericenter distance - eccentricity
          distribution plot for $n_{6}$. Two HMXB candidates (HMXBc)
          have been produced. {\it Right}: The same distribution plot
          for $n_{4}$. The circled points indicate HMXBc whose 
          mass transfer via RLOF may occur periodically due to their
          eccentric orbits. The elliptically-circled point indicates an
          HMXBc whose semimajor axis is smaller than $a_{{\rm RL}}$, so
          the mass transfer will be steady. There are a
          couple of binaries with pericenter distance smaller than
          $R_{{\rm RL}}$, but they are excluded because $M_{2}< 8 \Msol$.}
	\label{fig:8}
\end{figure*}

For the small scale case, we performed 86
runs of 5-body simulations for each of the number density values $n_6$ and $n_4$.
Due to the smaller initial separations of the stars, we run the simulations
for a shorter amount of time.

We run each simulation for a minimum of $1000\yr$, but
stop the run if a stable binary forms, and if
no further significant dynamical changes are observed.
If such a binary does not form, we run the simulations to a maximum duration of $5000\yr$.

In order to
properly compare the results from the $n_{6}$ calculations with those
from the $n_{4}$ calculations, we use the same initial conditions for
each set of runs.

In the $n_{6}$ calculations, the most common scenario
is that ${\rm S}_{1}$ always forms the most compact binary almost immediately,
while stellar scatterings are most common during 
the first few years to about 40 years.
Thereafter, a multiple system usually survives and stabilizes, while less massive stars are ejected.
A difference between the large-scale and the small-scale scenarios
is that, whereas most cases in the
large-scale calculations end up with one binary and unbound single stars
(in $15$ out of $20$ cases),
in the small-scale runs a multiple system such as triple or quartet
(rather than a simple binary) forms
(in 59 of 86 cases)
Less massive stars take some energy from the multiple system
and convert it into their kinetic energies
while causing the binary to harden.

As an example, Figure~\ref{fig:7} shows the
evolution in the semi-major axis of a compact binary in one of the
runs. One can easily notice the quick decrease in the semi-major axis
during violent stellar interactions (before 10~years), which clearly
indicates the hardening. After a couple of stars are cast away and the
multiple system is stabilized and isolated (after about 10 years),
dynamical friction plays the main role in the decrease of the
semi-major axis;
after this point, we do not expect further dramatic hardening of the binary.
In our 86 runs, the last surviving binary is typically of type ${\rm
  B}_{13}$ (pairing of the most massive and third most massive stars).
with total mass of $130\Msol$ and $\langle M_{{\rm
    S}_{3}}\rangle=11\Msol$. $\langle a_{t=5000~{\rm
    yr}}\rangle=1.87\AU$ and the minimum semi-major axis is
$0.2\AU$. The corresponding dynamical friction time scale (equation
\ref{eq11}) is $\sim 10^{13}$ yr. The ejected stars have speeds of
$(10-100)~c_{\rm s}$ , and $c_{\rm s}\sim 4~{\rm km/s}$.
  Six of the HMXB candidates (HMXBc hereafter) have been formed in all
  runs ($P_{{\rm HMXBc}}$ = 0.070). For later use, let us define
  $F_{\rm HMXB}$ as the number of HMXBc formed across all simulations,
  normalized by the total mass of the stars. Then $F_{{\rm HMXBc}} =
  4.6\times 10^{-4} \Msol^{-1}$ (This term will be used to estimate
  the X-ray luminosity and is one of the primary results of our
  study).

The outcomes of the $n_4$ simulations are quite similar: a triple or
higher multiple forms in 65 of 86 runs, and $\langle
  a_{t=5000~{\rm yr}}\rangle=1.42\AU$. Furthermore, due
  to the same number of HMXBc, $P_{{\rm HMXBc}}$ and $F_{{\rm HMXBc}}$
  are the same.  One notable difference is that for $n_4$, the
dynamical friction time scale is longer by 2 orders of magnitude
compared to $n_{6}$, because this quantity is inversely proportional
to the number density.

Interestingly, there are four runs (for each density value)
in which the most compact binary is eccentric,
and inside the requisite separation for RLOF at pericenter but
outside it at apocenter.
We consider only two of them as HMXBc
and rule out the other two binaries
since the mass of the more massive star  is smaller than $8 \Msol$ \citep{Heger2003}.
We present the distribution of eccentricity for
pericenter distance in Figure~\ref{fig:8}. In particular, the {\it left}
panel shows the distribution for $n_{6}$ calculations and the {\it
  right} panel for $n_{4}$. The circled point indicates HMXBc with
distance at pericenter shorter than the corresponding Roche-Lobe
radius.

How accretion proceeds in a highly eccentric binary system
under these conditions remains an unsettled issue to date, as studies
have claimed that the orbital semimajor axis and eccentricity can
either increase or decrease depending on the binary properties at
pericenter (\citealt{Sepinsky2007} and \citealt{Sepinsky2009}). If the
RLOF does induce circularization, then accretion proceeds normally
(i.e. steadily). However, if the RLOF instead 
increases the eccentricity of the system, then, whether accretion can proceed steadily
rather than intermittently will depend on the relative timescale between
the disk lifetime $\tau_{\rm disk}$ (on the order of the viscous timescale),
and the orbital period of the binary $t_{\rm orb}$.
We computed these timescales for all the eccentric binaries
in our simulations (for which the Roche-Lobe radius straddles the pericenter and apocenter),
and found that $\tau_{\rm disk} > t_{\rm orb}$ in all cases but one.

This implies that the fraction of binaries whose eccentricities cause intermittent RLOF
is small, and that as a global average, RLOF is steady to a good approximation.

   \begin{figure*}
   	\centering
   	{\includegraphics[width=8.5cm]{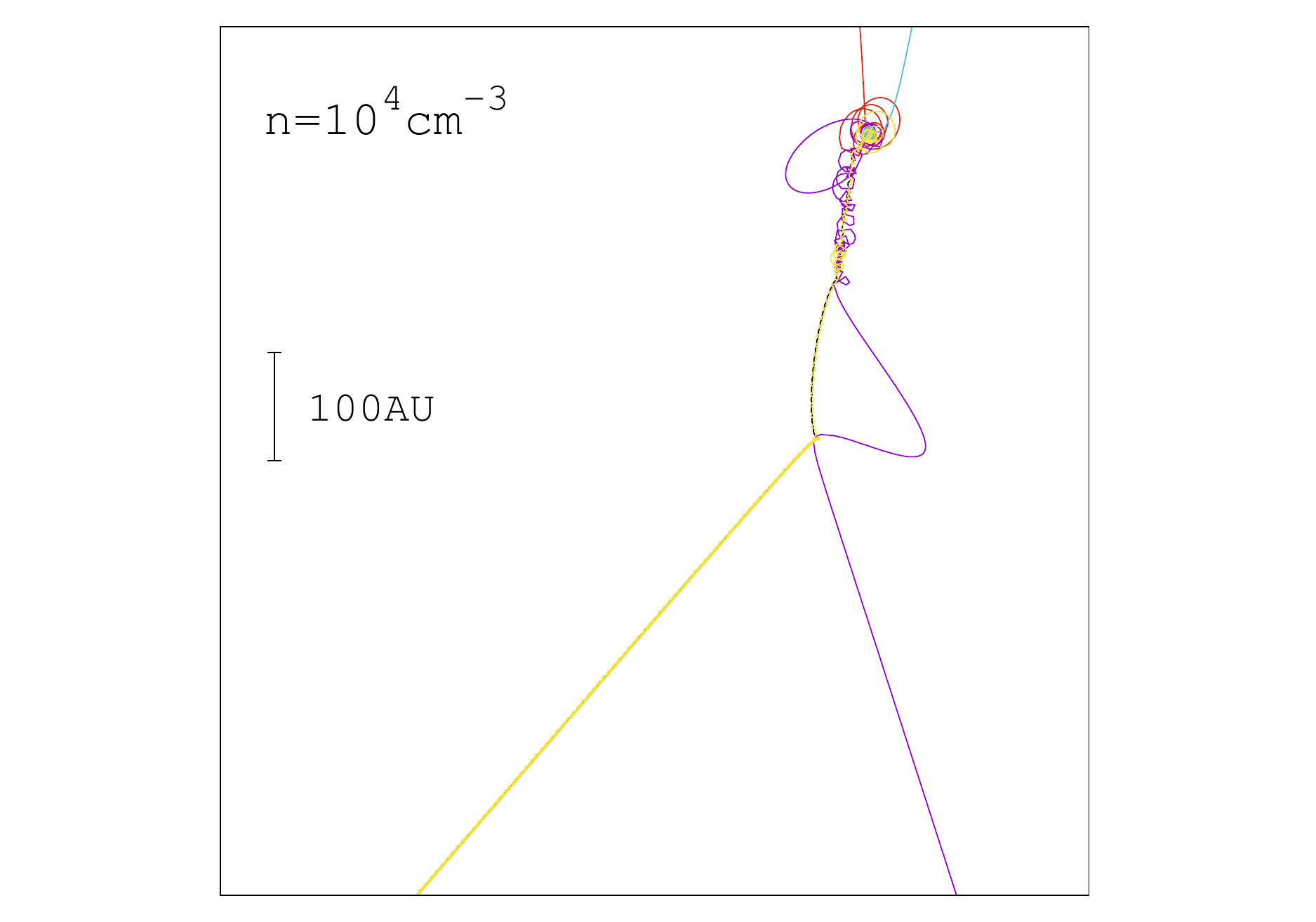}}
   	{\includegraphics[width=8.5cm]{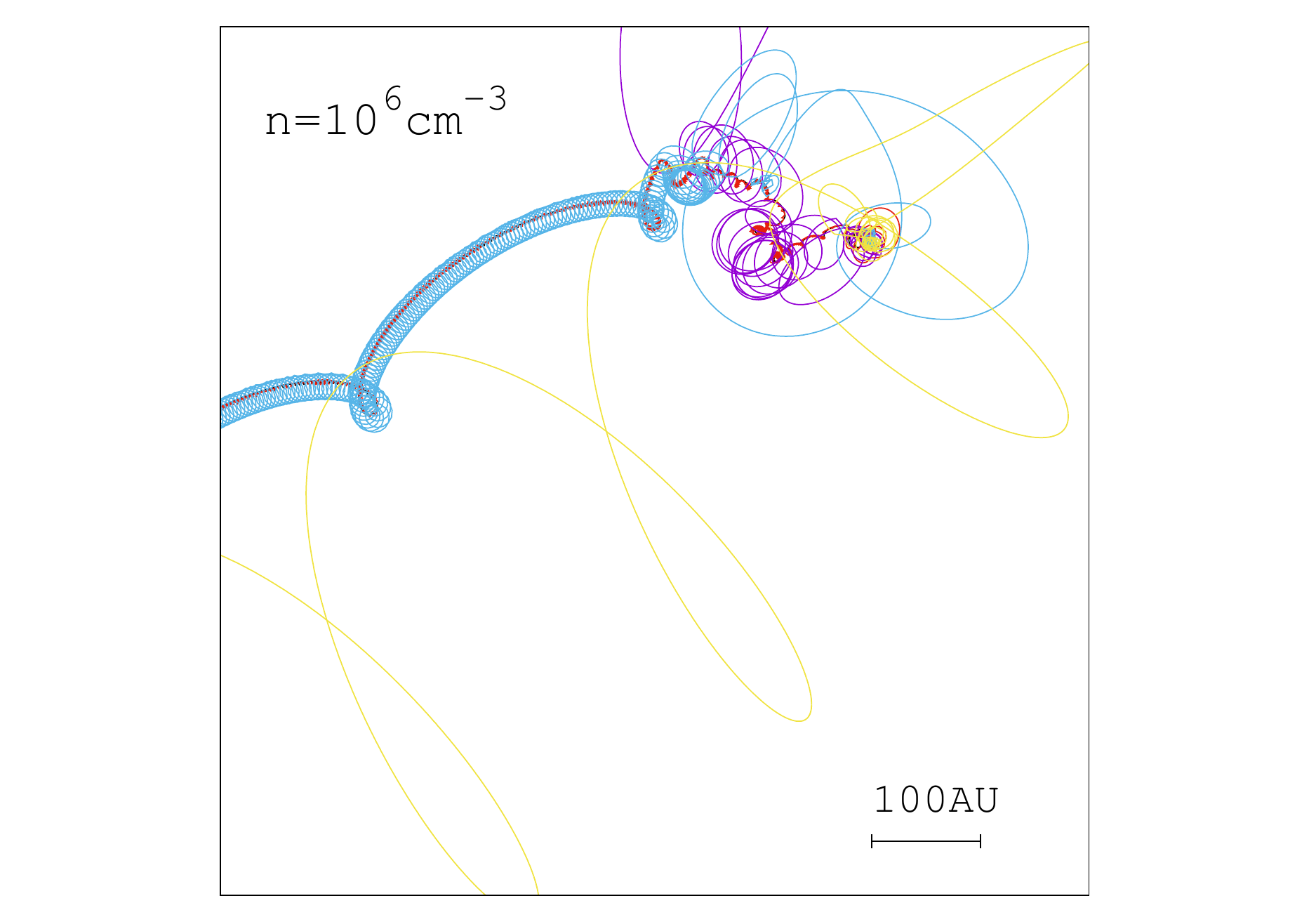}}		
   	\caption{
	Trajectories up to $t = 2000~{\rm yr}$ for 5-body simulations with identical
initial conditions but different densities: $n_{4}$ (left panel) and $n_{6}$ (right panel).
Despite having the same starting point, the trajectories
of the 5 stars evolve quite distinctly in the two backgrounds due to the different magnitudes of
dynamical friction. 
 The two simulations for different bound systems: a binary
          for $n_{4}$ (yellow line) and a quartet for $n_{6}$
          (red+blue+yellow lines). }
   	\label{fig:4}
   \end{figure*}
 
 \begin{figure*}
 	\centering
 	{\includegraphics[width=8.5cm]{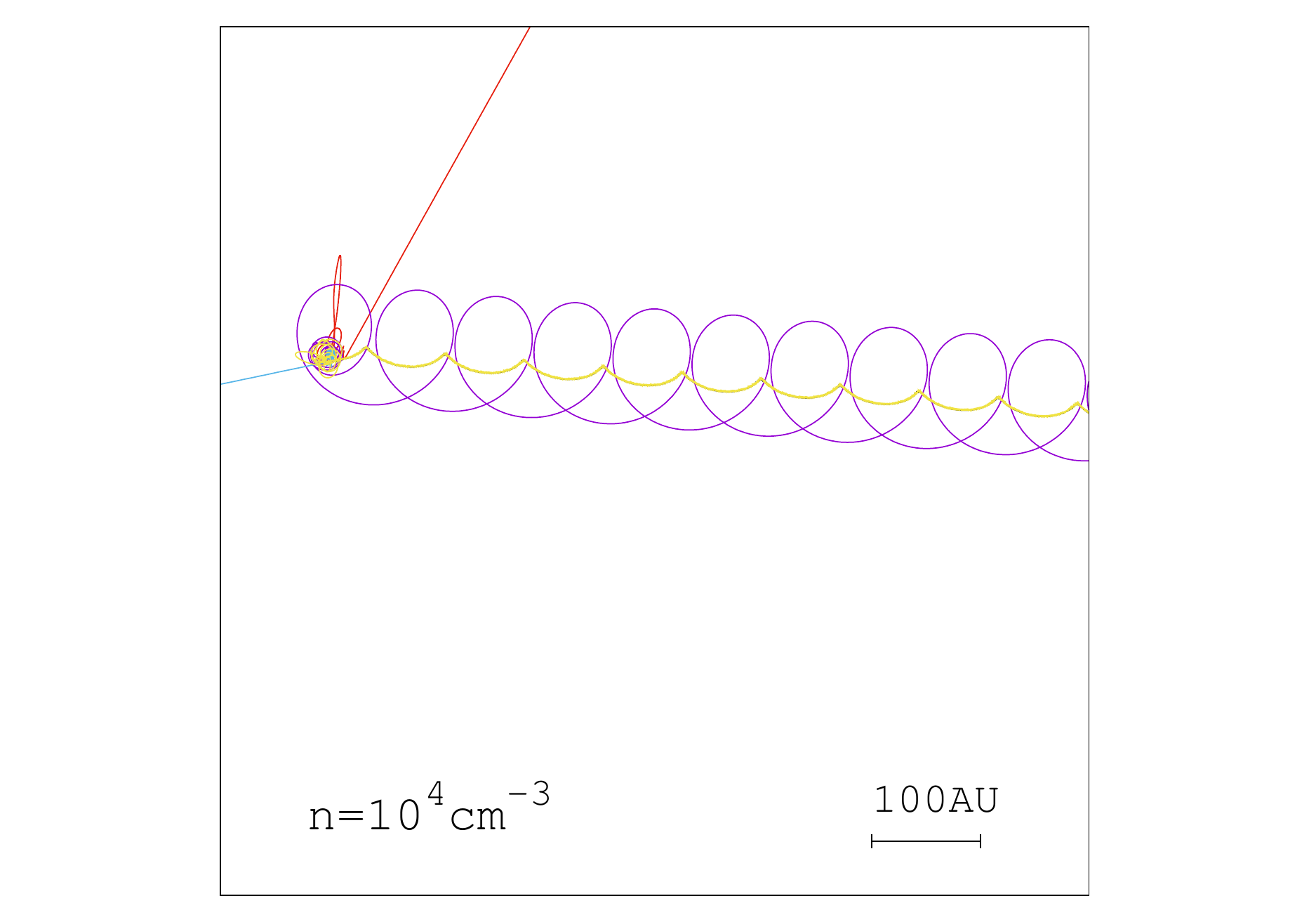}}
 	{\includegraphics[width=8.5cm]{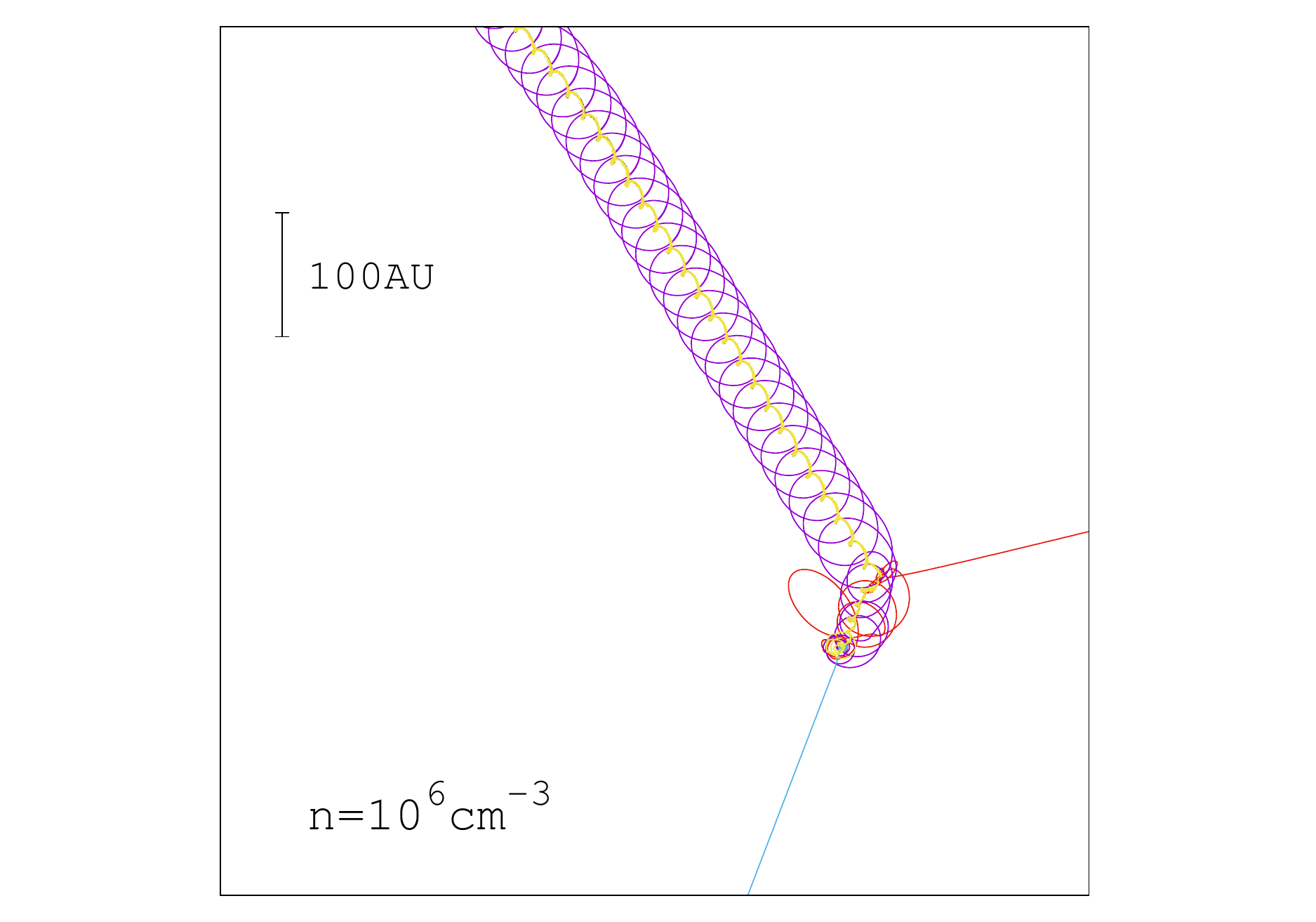}}		
 	\caption{
	Trajectories of two 5-body simulations with identical initial conditions
	but different ambient gas densities.
	The simulations have formed triples with the identical stars (yellow+purple lines; contrast with \ref{fig:4})
	at $t = 1000\yr$, despite differences in the trajectories of each star and the center of mass.
	}
 	\label{fig:3}
 \end{figure*}
 
\begin{figure}
	\centering
	{\includegraphics[width=8.5cm]{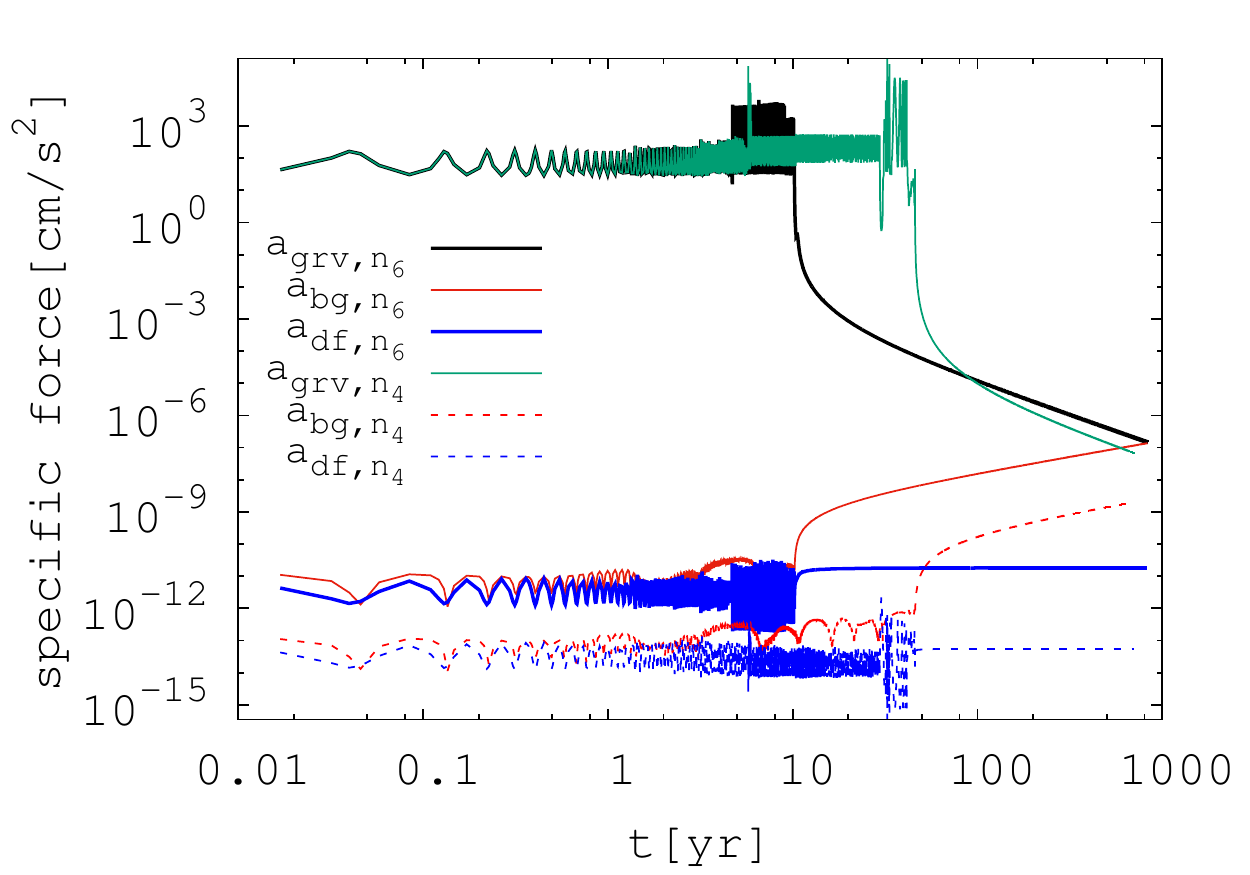}}
	\caption{Changes with time in the
          three forces per unit mass (gravity from other stars,
          dynamical friction and background gravity) acting on
          star$_{4}$ (the light blue color line in Figure~\ref{fig:3}).
          We run two simulations with identical initial conditions
          but different ambient gas densities $n_4$ and $n_6$.
          During the early phases
          of the interaction, the gravitational force
          ($a_{{\rm gr}}$) dominates by several orders of magnitude over dynamical
          friction ($a_{{\rm df}}$) and the background force
          ($a_{{\rm bg}}$).
          Note that $a_{{\rm gr},n_{6}}$ (black line) and $a_{{\rm gr},n_{4}}$
          (green line)  synchronize at early times because
          the motion of the star is very close to Keplerian and $r
          \sim{1}/{v^{2}}$.
          After the star is ejected at 10~yr
          for $n_{6}$ (later for $n_{4}$), $a_{{\rm gr}}$
          monotonically decreases and $a_{{\rm bg}}$ increases
          ($a_{{\rm bg}} \sim r$ where $r$ is the distance from the
          center of the halo). During the same time, $a_{{\rm df}}$ barely changes
          since the star moves in the supersonic regime and the speed
          decreases very slowly.  }
	\label{fig:2}
\end{figure}

The magnitude of the gas density considerably influences the characteristics of 
the dynamical interactions.
While on the one hand $40\%$ of the simulations end up forming the same triples
for both density values, on the other hand their trajectories and the center of mass movements
differ significantly.
Figure~\ref{fig:4} shows two sample trajectories 
of 5 stars with low number density (left panel) and high number density
(right panel) after 1000 yr. They were given identical initial conditions
for the run, but their trajectories have developed
differently. In Figure~\ref{fig:3}, at 800 yr, even though the same stars
form a triple and the same star is ejected for both number densities, 
their trajectories are clearly different.

As can be seen in
Figure~\ref{fig:2}, which shows the change of forces per unit mass exerted
on a typical kicked-off star, $a_{{\rm df}}$ and $a_{{\rm bg}}$ are
negligible compared to $a_{{\rm gr}}$ before about 5 years.

Note
that $a_{{\rm df}}$ and $a_{{\rm bg}}$ are synchronized at early times 
because the motions of the stars are close to the Keplerian motion,
$r\sim1/v^{2}$. The star in Figure~\ref{fig:2} is ejected at $t \approx 10~
{\rm yr}$ for the high number density (and at $60\lesssim t 
\lesssim 80~{\rm yr}$ for the low number density). This can be understood from the fact that
$a_{{\rm gr}}$ monotonically decreases and $a_{{\rm bg}}$ increases
($a_{{\rm bg}}\sim r$ where $r$ is the distance from the center of
mass). At the same time, $a_{\rm df}$ barely changes since the star
moves in the supersonic regime and the speed decreases slowly.

There is no noticeable difference in the overall results 
between the $n_4$ and $n_6$ runs---quantities such as 
$\langle a_{t=5000~{\rm yr}}\rangle$, the dynamical friction time scale
$\tau_{\rm df}$, the total mass of the most compact binary (or
which star forms the compact binary with ${\rm S}_{1}$) as summarized
in Table \ref{tab:tab2}. For both number densities,  the
companion star of the binary is typically the third most massive star
${\rm S}_{3}$.

To sum up, we find that $P_{\rm HMXBc}\sim 7\%$ of our simulations form HMXBc, regardless
of the gas density value.
Normalized to the total stellar mass in the simulations, the number
of HMXBc formed per stellar mass is
$F_{\rm HMXB}\approx 4.6\times 10^{-4}\Msol^{-1}$.

\subsection{10-body simulations}

\label{subsec:10-body}

We now explore several different configurations for the star group,
and run several sets of simulations with 10 stars (instead of 5).
These are: (1) 10-body version of the small scale calculation presented above;
(2) head-on crash of two star groups containing 5 stars each;
(3) a close encounter and subsequent inspiral and merger of two star groups
containing 5 stars each. 
The latter two scenarios are motivated by the fact that the merger timescales 
and mass accretion timescales of Pop~III host haloes, as well as the lifetimes
of the massive Pop~III stars themselves, are of the same order, $\sim 10\Myr$.
This suggests that merging haloes will be continuously forming new stars
(perhaps Pop~II instead of Pop~III) as they merge with other haloes,
and that close interactions and mergers of nascent star groups may be relatively common.
We generate stellar masses in the same way as for the 5-body case,
but with a larger value for the parameter $M_{\rm max}=300\Msol$.
We have run each simulation for 500 yr. The
results of these simulations are summarized in Table \ref{tab:tab3}.
We briefly discuss each one, as follows.

\begin{table}
	\centering
 \setlength\extrarowheight{2pt}
	\begin{tabulary}{0.6\textwidth}{c c c c c c}
		\hline
		 Scenario & 1 & 2 & 3 & 4A & 4B \\
		\hline
		runs 													      & 54	   	  &	 30	  	& 30	 	 & 	30	  	&	30	\\
		$\langle a_{t=500\yr}\rangle$   $[\AU]$     & 1.0 	  	&  1.1	  & 0.90	  &	1.8	  	 & 	 1.6 \\
		$\langle a_{\rm{B}_{12}}\rangle$ $[\AU]$  & 1.6	  	  &  1.5	 &	1.8	  	&   2.4	 	&    2.0	  \\
		$\langle a_{\rm{rest}}\rangle$  $[\AU]$ 	  &  0.72   &  	0.15  & 0.42	 & 1.2		&	0.23  \\
		$P_{\rm{HMXBc}}$  									&  0.33	   &  0.33	&  0.27		&	0.13  & 	0.27  \\
		$F_{\rm{HMXBc}}$ [$10^{-4}\Msol^{-1}$] &    15	   &  11	  &  9.0	  &	4.2      & 	8.4  \\
		\hline
	\end{tabulary}
		\caption{Summary of the results from 10-body simulations. We have
		considered five
		different scenarios.
		Scenario 1 simulated the 10-body version of the 5-body calculation,
		i.e. an isolated star group.
		The other scenarios all involve collisions of two 5-body star groups.
		Scenario 2 is a head-on collision of two coplanar, co-rotating star groups.
		Scenario 3 also collided two groups of stars with co-rotating orbital planes,
		but with an impact parameter comparable to
		the sizes of the groups, which results in an inspiral and eventual merger.
		Scenarios 4A and 4B are similar to scenario 3, except that
		the orbital planes of the colliding star groups had
		mutual inclinations of $45^\circ$ and $135^\circ$, respectively.
		}
	\label{tab:tab3}
\end{table}

\subsubsection{Scenario 1: 10-body group in isolation}
\label{subsubsec:Scenario1}
We set up the simulations as in the 5-body calculations, but with 10 stars.

We find that $\langle a_{t=500{\rm yr}}\rangle=1.0~{\rm
  AU}$. Interestingly, there are 18 out of 54 cases in which
$a<a_{{\rm RL}}$. There is a large difference in scale between
$a_{{\rm B}_{12}}$ and $a_{{\rm rest}}$, where $a_{{\rm B}_{12}}$ is
the semi-major axis of ${\rm B}_{12}$ (binary made up of the two most
massive stars) and $a_{{\rm rest}}$ is the semi-major axis of the
binary stars other than ${\rm S}_{1}$ and ${\rm S}_{2}$. 
  This is a common feature of the 10-body simulations:
  they often end up with triples whose inner
  binary is $B_{\rm rest}$ while the outer binary is $B_{12}$. Our
simulations yield $\langle a_{{\rm rest}}\rangle =0.72~{\rm AU}$ and
$\langle a_{{\rm B}_{12}}\rangle=1.6~{\rm AU}$. Also note that, in 14
out of 18 HMXBc, the binary is ${\rm B}_{{\rm rest}}$ (i.e. it is not
made up of the two most massive stars).

Since the compact binaries in these simulations form quickly
and we only follow them for 500 years, 
it is technically possible that they will be disrupted before one of the stars
turns into a BH. However, our simulations for the 5-body scenario
showed that compact, quasi-steady binaries are unlikely to be disrupted,
and for practical purposes we extrapolate this qualitative result to the 10-body case.

We find that a HMXBc forms in a larger fraction of these simulations
than in the 5-body case, $P_{\rm HMXBc}=0.33$,
for the obvious reason that there are more stars. 
Per unit stellar mass in the simulations, the number of HMXB candidates
is $F_{\rm HMXB}=1.5\times 10^{-3}\Msol^{-1}$,
which is a factor $\approx 3$ higher than we found for the 5-body case.
\begin{figure*}
	\centering
	{\includegraphics[width=5.8cm]{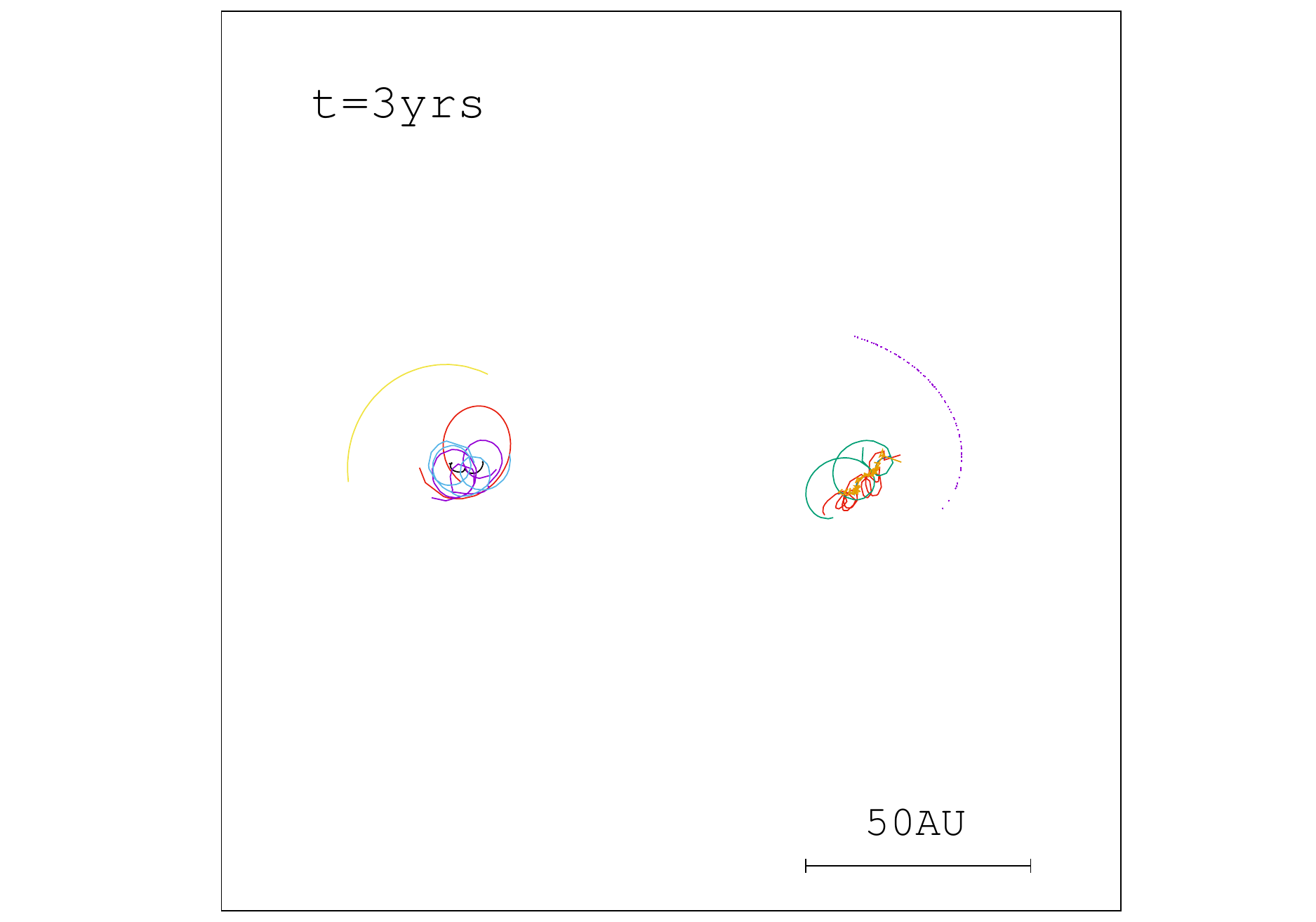}}
	{\includegraphics[width=5.8cm]{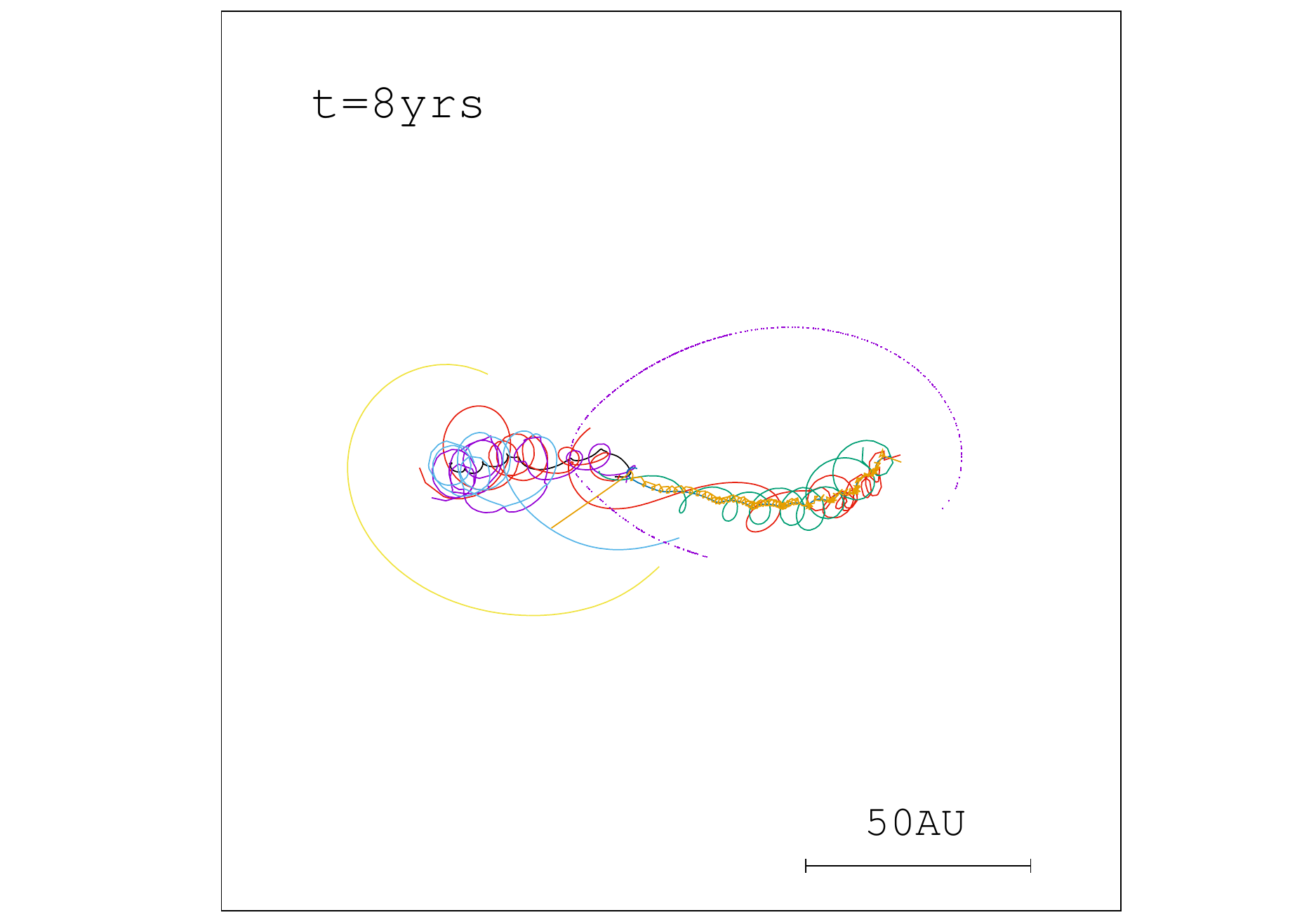}}
	{\includegraphics[width=5.8cm]{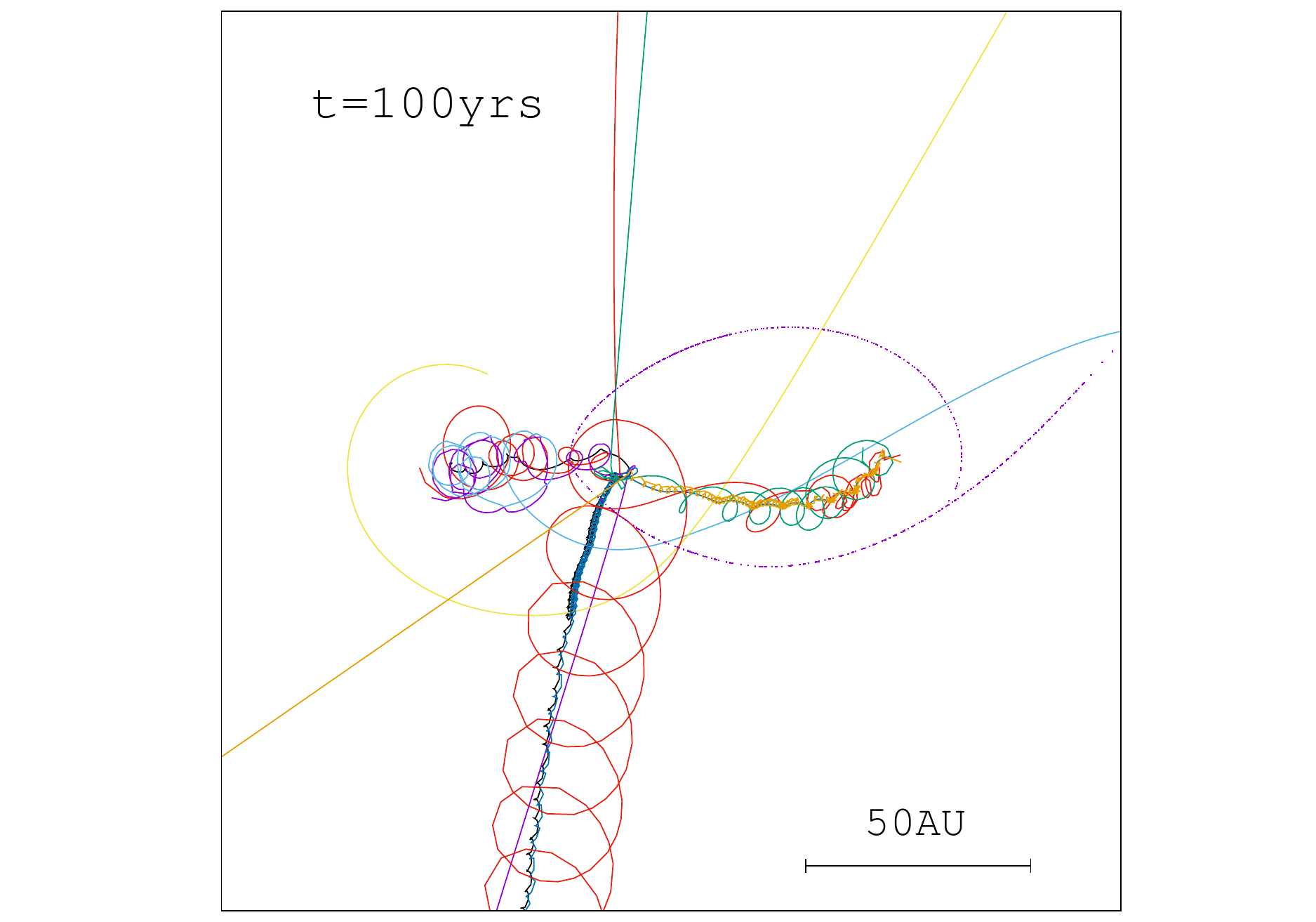}}				
	\caption{ 10-body head-collision (scenario 2). In
		these sample trajectories, five stars in each group are in nearly
		Keplerian motion. The groups move towards each other
		with a relative velocity of $\sim c_{\rm s}$ (left
		panel). At $t\approx 8$~yr (middle panel), the two haloes
		begin to merge. During the merger, all ten stars undergo
		stellar scatterings (right panel). One triple (black+dark blue+red lines)
		has been formed and it is moving in -y direction. In these
		head-on collisions, it is likely for multiple systems that
		existed before merging to be broken, while a few new
		multiple systems are formed.}
	\label{fig:10}
\end{figure*}

\begin{figure*}
	\centering
	{\includegraphics[width=5.8cm]{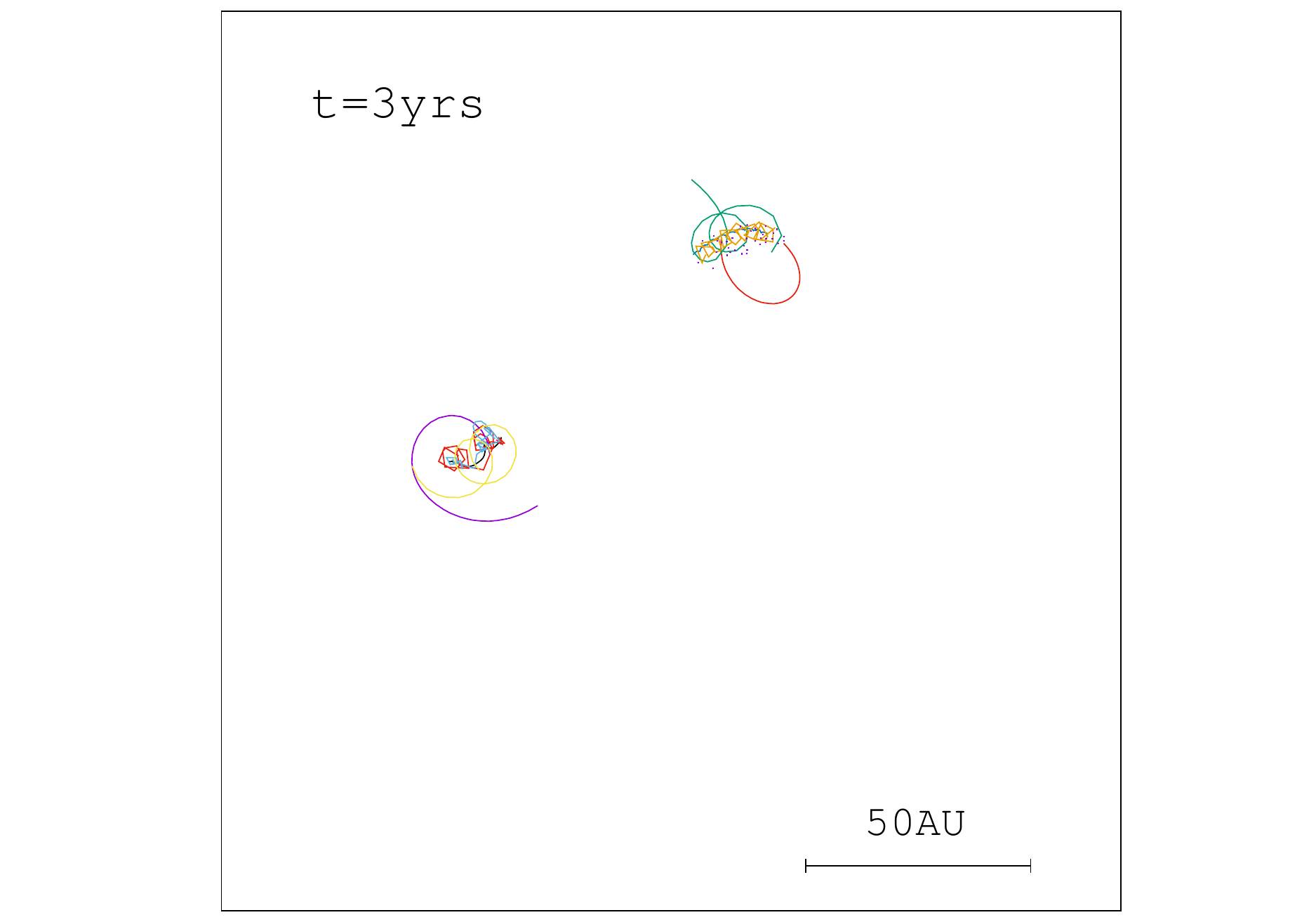}}
	{\includegraphics[width=5.8cm]{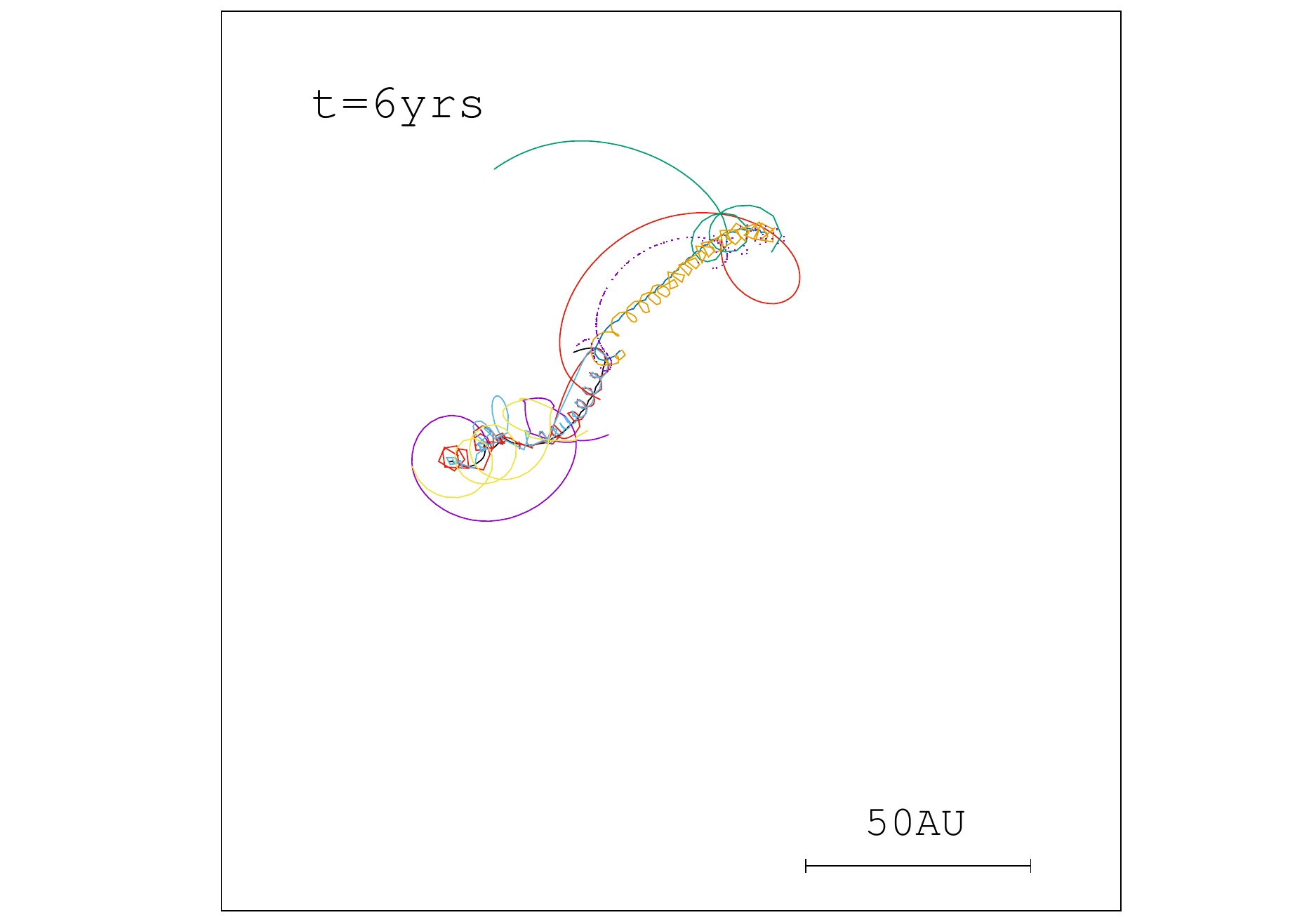}}
	{\includegraphics[width=5.8cm]{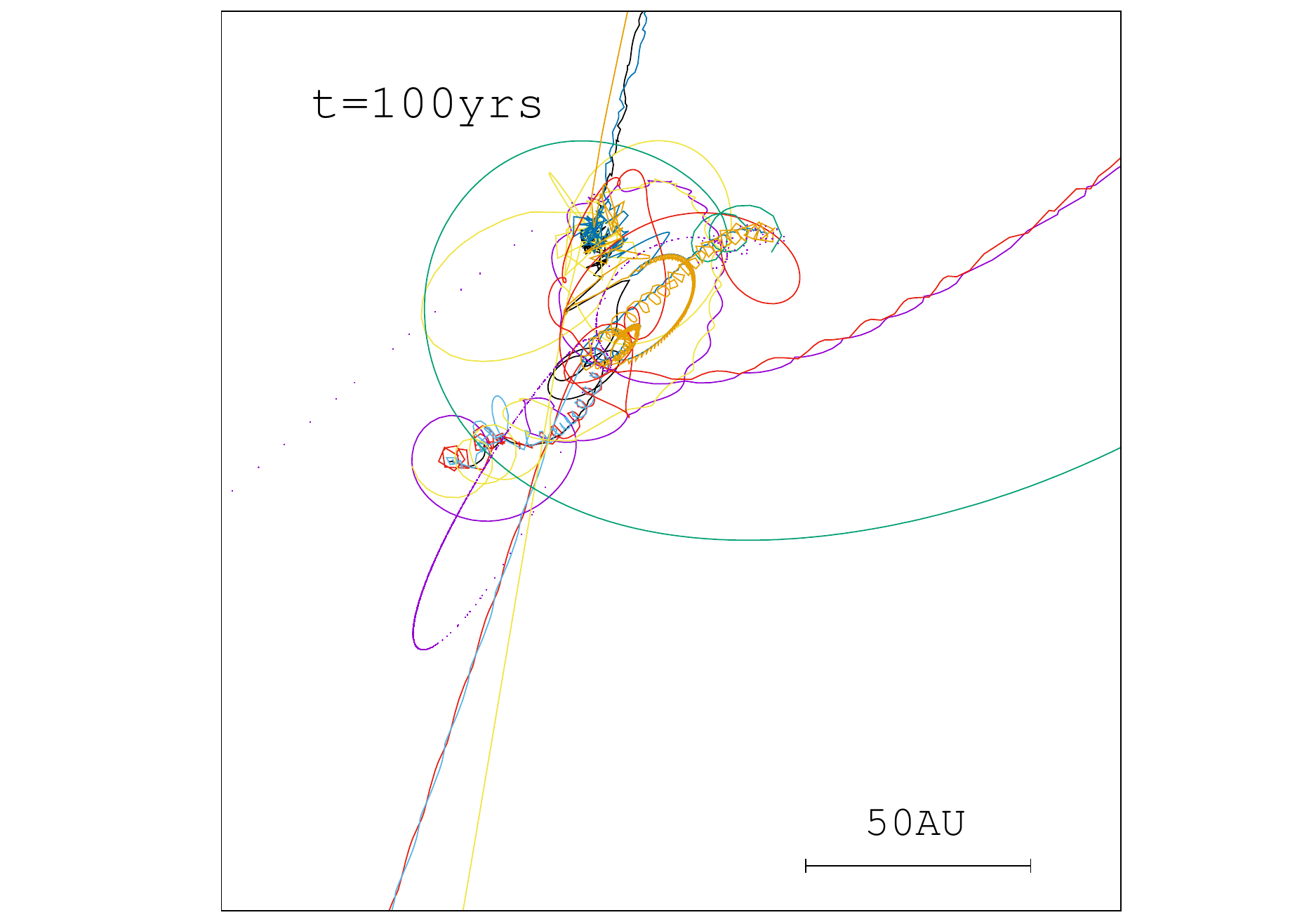}}				
	\caption{10-body spirally merging case
		(scenario~3). As with scenario 2, two groups of 5 stars
		are set on a collision cours. However, in this scenario the impact parameter is
		$\sim 2-3$ times the size of groups, whereas in
		scenario~2 it is set to be zero. The groups are
		approaching each other at the relative velocity of $c_{\rm s}$
		(left panel). At $t=8$~yr, the groups are about to
		merge (middle panel). After some time (right panel),
		two binaries are ejected (red and purple lines
		in the $x$-direction and black and blue lines in the $y$-direction). In this
		sample case, the most compact binary is the one ejected in
		the y direction (black+blue lines) with $a=0.8$~AU at
		$t=100$~yr.  A difference between this scenario and the head-on collision
		is that compact systems are more likely to survive
		the merger.}
	\label{fig:11}
\end{figure*}
\subsubsection{Scenario 2: Collision between two 5-star groups -- head-on collision}
\label{subsubsec:Scenario2}

Two groups of 5 stars are set up with random initial conditions,
in the same manner as for the previous simulations of 5-body groups.
The two groups are then arranged to collide head-on, as follows:
they are placed at a separation of two to three
times their sizes and their disks are aligned so that the mutual inclination is zero. 
The initial relative speed of the groups is roughly
the speed of sound and the center of mass of one group is set to move
directly toward the center of mass of the other group.

We find that prior to colliding, each group forms a compact binary of
type ${\rm B}_{12}$ (the most and second-most massive
star). When the two groups collide, those two binaries
  that existed before the collision were broken and the two most
  massive stars of each group form a new compact binary with high
  chances.

The
average $\langle a_{t=500~{\rm yr}}\rangle$ is $1.1~{\rm AU}$, but
$\langle a_{{\rm B}_{12}}\rangle=1.5~{\rm AU}$ and $\langle
a_{{\rm rest}}\rangle =0.15~{\rm AU}$, meaning that the most compact
binaries are not formed from the most massive stars.
The shorter average separations may be a result of a larger
number of early 3-body scatterings, which act to harden 
the group as a whole.

HMXBc form in 10 out of 30 runs, and they are not of type ${\rm B}_{12}$.
However, we find a rate of HMXB formation per stellar mass
$F_{\rm HMXB}=1.1\times 10^{-3}\Msol^{-1}$;
this is higher than in the 5-body case and comparable to Scenario 1 above.
Sample trajectories 
for one of the simulations of scenario 2 are depicted in Figure~\ref{fig:10}.

\subsubsection{Scenario 3: Collision between two 5-star groups -- spirally merging case}
\label{subsubsec:Scenario3}

We have used the same input parameters for the two groups as in
Scenario 2, except that we now set the impact parameter to be of order
the size of the group, whereas it was set to zero in Scenario 2.

The groups are given opposite velocities of $\sim c_{\rm s}$ along the $x$-direction,
and are offset by a displacement along the $y$-direction
that is $\sim 2 - 3$ times the typical size of the star group 
($\sim 20~{\rm AU}$) so that they merge with a spiral motion.

We find that  $\langle a_{t=500~{\rm yr}}\rangle = 0.90\AU$, $\langle
a_{{\rm B}_{12}}\rangle=1.8\AU$ and $\langle a_{{\rm rest}}\rangle
=0.42\AU$.

The average separation lies between what we find in Scenarios 1 and 2.
This can be interpreted as being due to the fact that these simulations
(in which the two groups merge gradually via inspiral)
have more close 3-body interactions than in Scenario 1 (in which
10 stars in quasi-Keplerian orbits evolve in isolation)
but fewer such interactions than in Scenario 2
(in which the two groups merge head-on).

HMXBs form in 8 out of 30 runs and, as with Scenarios 1 and 2,
none of the HMXBs are made up of the two most massive stars.
We find a similar HMXB formation rate per stellar mass,
$F_{\rm HMXB}=9.0\times 10^{-4}\Msol^{-1}$.
Sample trajectories from one of the
simulations for Scenario 3 are shown in Figure~\ref{fig:11}.

\begin{figure*}
	\centering
	{\includegraphics[width=5.8cm]{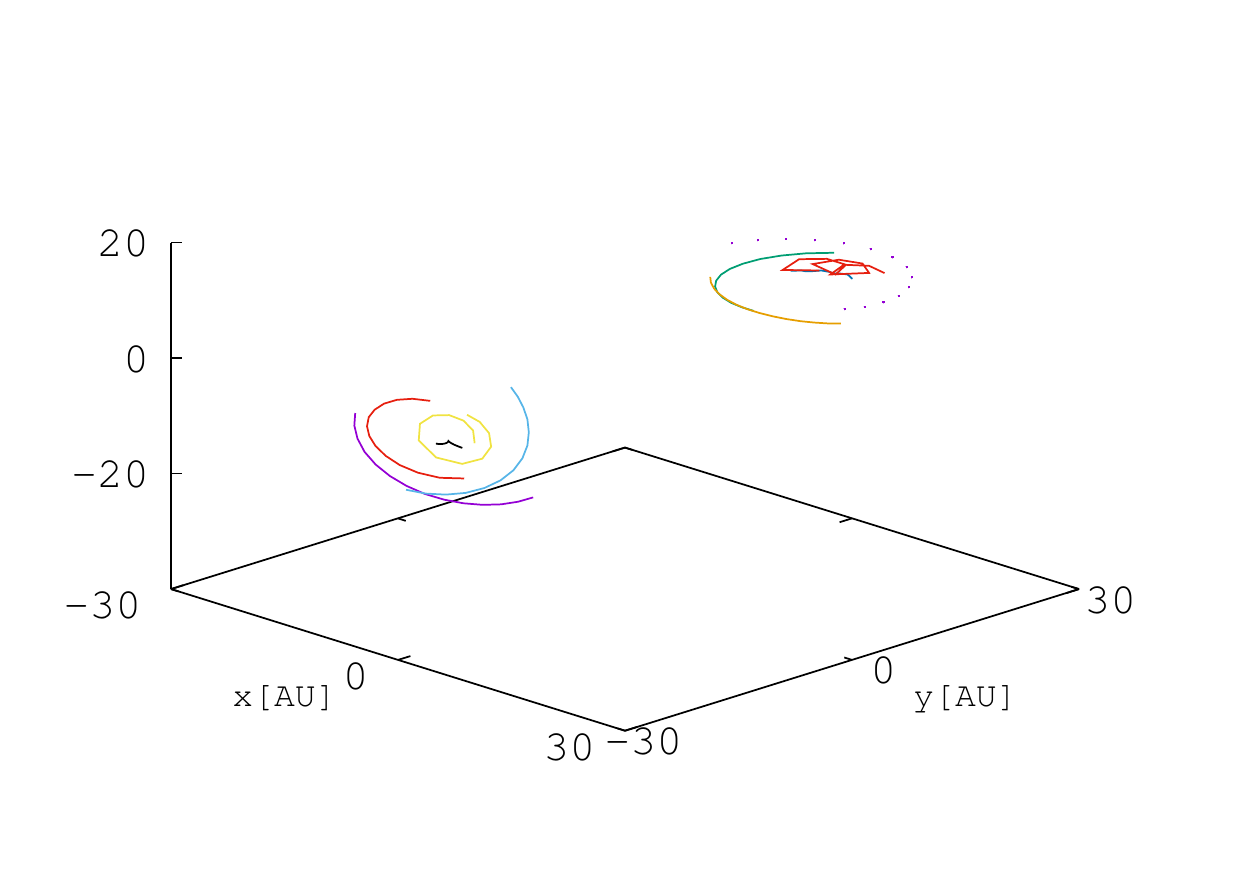}}
	{\includegraphics[width=5.8cm]{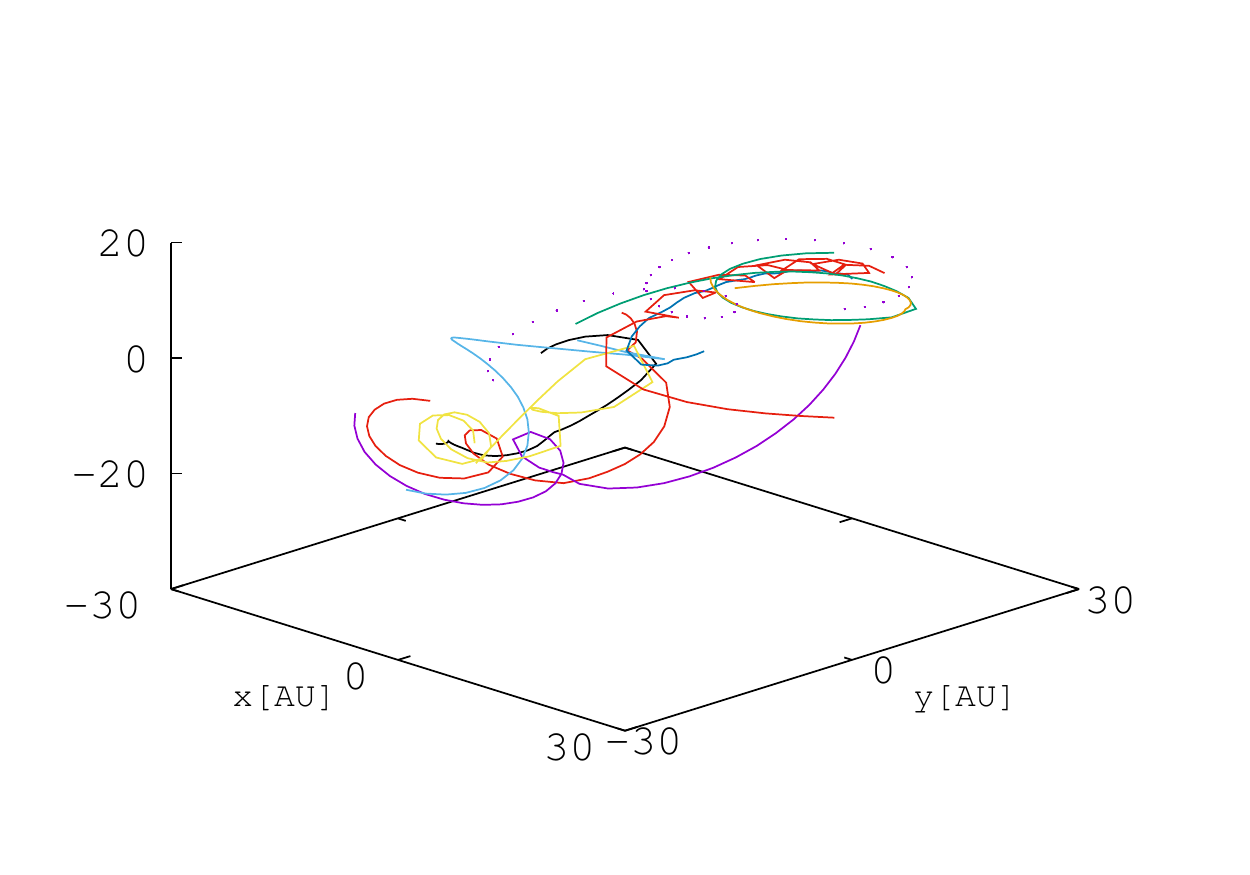}}
	{\includegraphics[width=5.8cm]{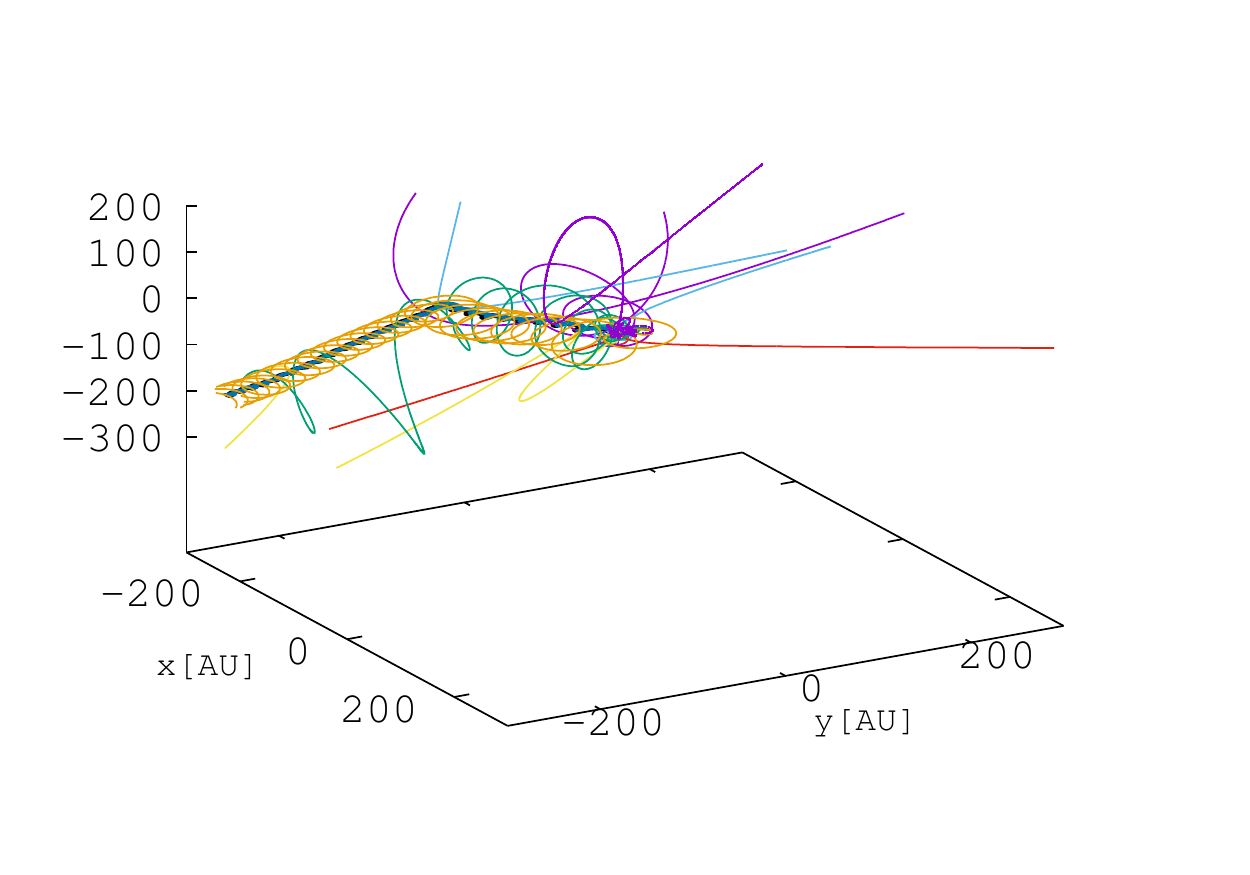}}				
	\caption{Trajectories for scenario 4A, a collision between two 5-body groups, each
		on quasi-Keplerian orbits, but with the orbital plane
		of the two groups tilted with respect to each other
		at an inclination  $i= 45^{\circ}$.
		The setup is the same as scenario 3,
		except for the mutual inclination of the orbital planes of the colliding
		star groups.
		The left panel shows the two groups on a collision course.
		At $t=4\yr$, the halos are about to
		merge (middle panel). After $t=10\yr$ (right panel),
		a triple having the most compact binary are ejected [black and blue lines (inner binary) and brown line].}
	\label{fig:12}
\end{figure*}

\begin{figure*}
	\centering
	{\includegraphics[width=5.8cm]{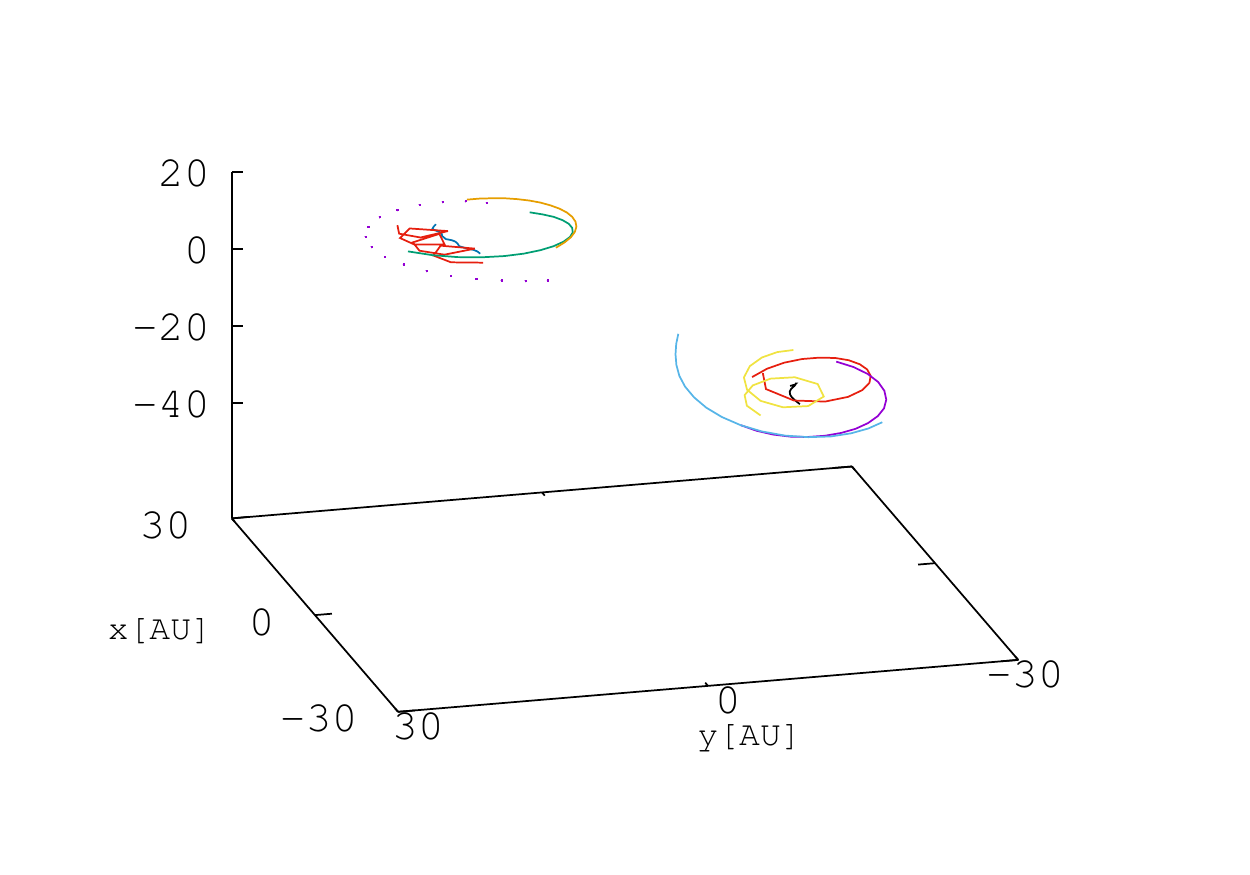}}
	{\includegraphics[width=5.8cm]{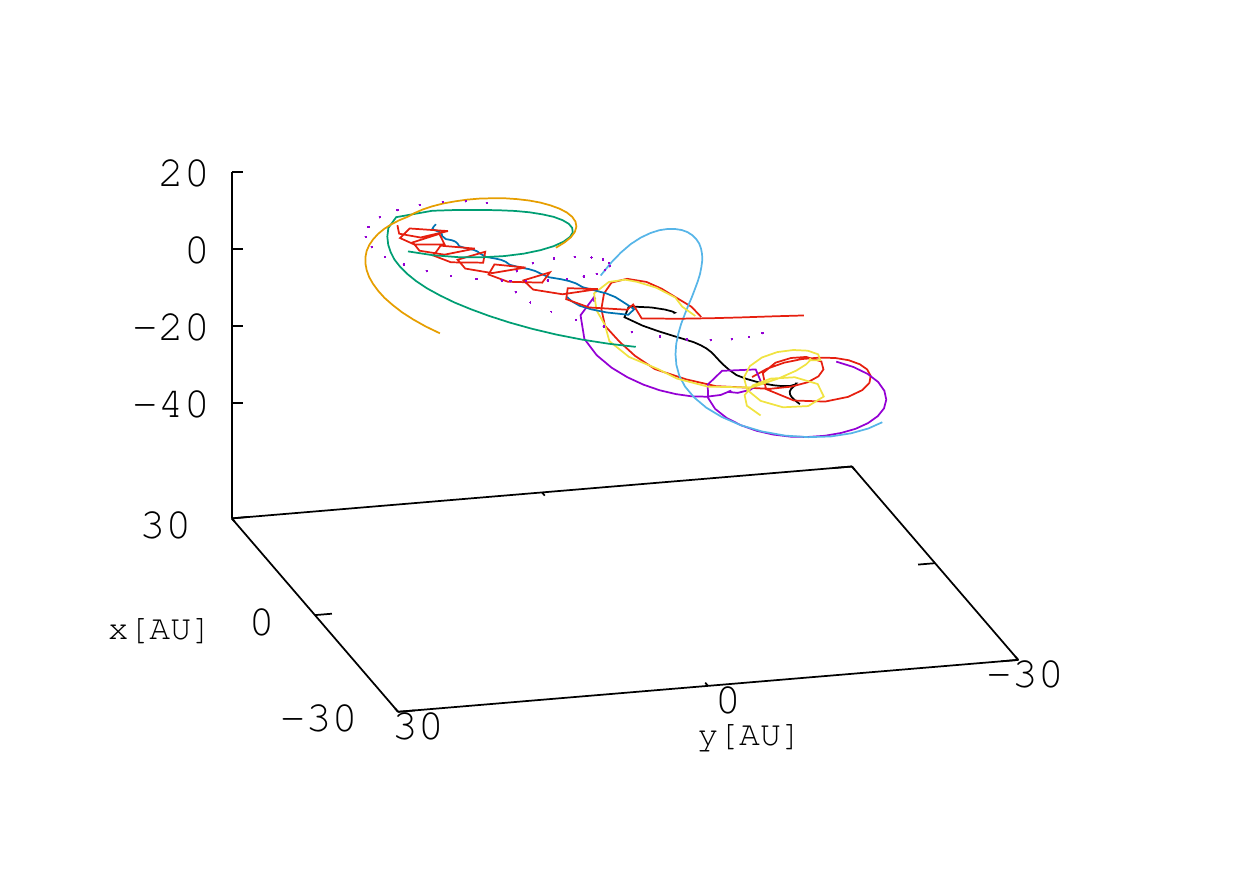}}
	{\includegraphics[width=5.8cm]{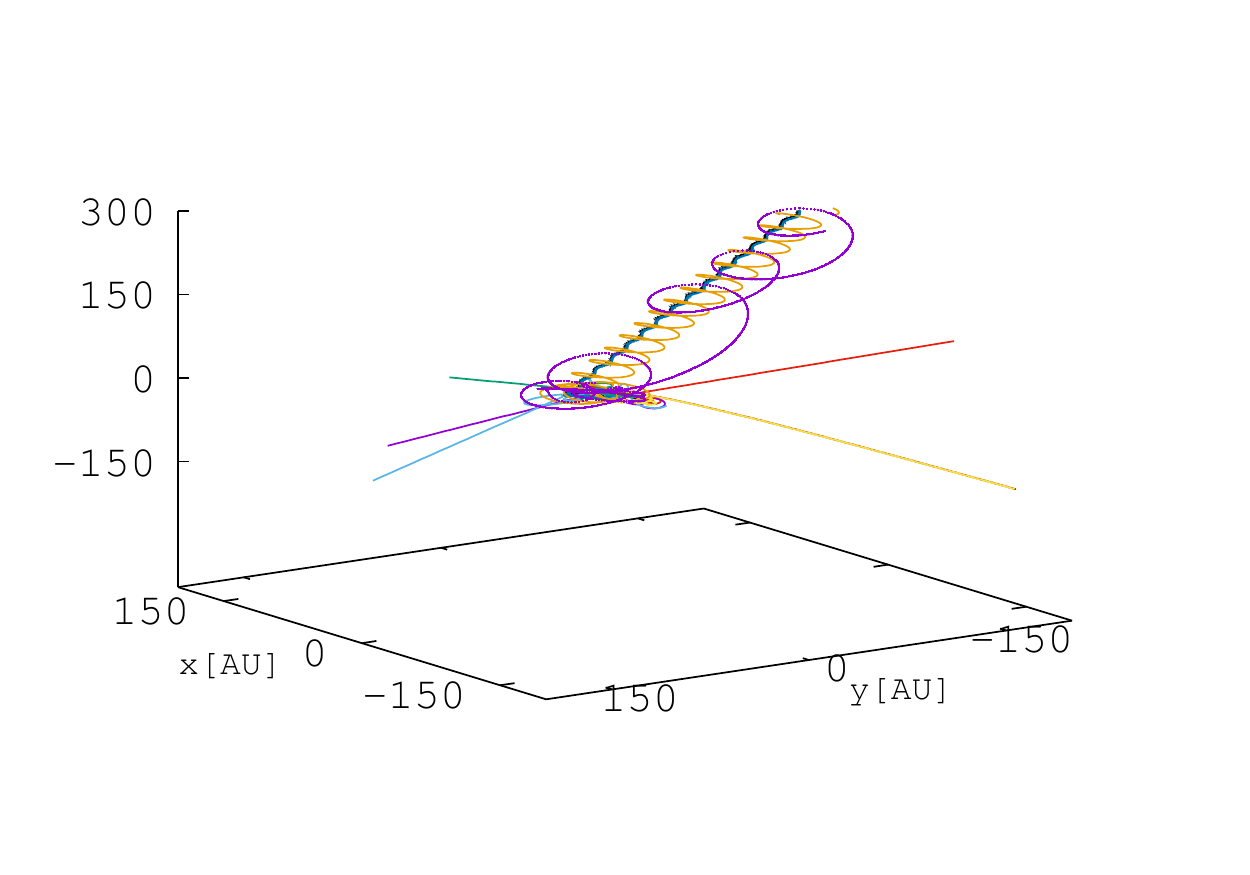}}				
	\caption{Trajectories for scenario 4B---the same as scenario 4A (Figure~\ref{fig:12}),
		but with the two groups nearly counter-rotating with respect to each other
		($i=135^{\circ}$). At $t=4\yr$,
		the groups are about to merge (middle panel) and after $t=10\yr$ (right panel),
		a quartet containing the most compact binary is ejected toward the upper right
		of the panel [black+blue lines (inner binary), brown and thick purple lines].}
	\label{fig:13}
\end{figure*}
\subsubsection{Scenarios 4A and 4B: Collision between two 5-body groups -- spirally merging case with inclinations of 45 degrees and 135 degrees}
\label{subsubsec:Scenario4}
In these two scenarios, 
we again set two groups of five stars each on a collision course.
The difference is that the orbital plane of one star group is tilted,
so that the two stellar disks have a mutual inclination $i$.
We set the inclination at $i=45^\circ$ (nearly co-rotating) for scenario 4A,
and $i=135^\circ$ (nearly counter-rotating) for scenario 4B.
The groups are initially placed at a separation of two to three
times their sizes, and set in motion at the same speeds
as for the inspiral case (scenario 3).

We find that $\langle a_{t=500\yr}\rangle = 1.8\AU$,
$\langle a_{\rm{B}_{12}}\rangle=2.4\AU$,
and $\langle a_{\rm{rest}}\rangle=1.2\AU$ for scenario 4A.
In scenario 4B, we find
$\langle a_{t=500\yr}\rangle =1.6\AU$,
$\langle a_{\rm{B}_{12}}\rangle=2.0\AU$,
and $\langle a_{\rm{rest}}\rangle=0.23\AU$.
In $70\%$ of the runs for both scenarios, the two most massive
stars form the most compact binary.
Stellar binaries end up with somewhat closer separations in the nearly counter-rotating case,
due to the fact that the net angular momentum of the merged star group is smaller.
Indeed, we find a total of four HMXBc across all the $i=45^\circ$ simulations,
and eight in a same number of $i=135^\circ$ simulations.

For the same reason, we find a larger fraction of
HMXB candidates per stellar mass simulated in the nearly counter-rotating case
($F_{\rm HMXB}=8.4\times 10^{-4}\Msol$)
compared to the nearly co-rotating case ($F_{\rm HMXB}=4.2\times 10^{-4}\Msol$).
The overall formation rate of HMXB candidates is lower for both cases than the
cases in which all the stellar orbits were nearly coplanar (Scenarios 1, 2 and 3),
plausibly due to the additional degree of freedom in the stellar orbits.
Still, the value of $F_{\rm HMXB}$ is within a factor of a few for all of our simulations.

Sample trajectories from runs for scenarios 4A and 4B are depicted in
Figure~\ref{fig:12} and Figure~\ref{fig:13}, respectively.

\section{Discussion}
\label{sec:discussion}

\subsection{Binary evolution and formation of HMXB candidates}
\label{subsec:Binaryevolution}

Our large-scale simulations show that the if Pop~III stars form hundreds of AU apart,
as in the simulations of \citet{stacy13}, then the timescales required
to make HMXBs via stellar scatterings are simply too long.
On the other hand, if protostellar clouds fragment and form stars in close
groups on scales of $\simgt 10\AU$, as in \citet{greif11}, then a small fraction
of groups can form HMXBs. We briefly discuss the dynamics of
HMXB formation in our simulations, then move on to discuss the
astrophysical implications of our findings.

The simulations indicate that, as expected,
scatterings play a major role in making a compact binary.
The background potential and the dynamical friction play a secondary role,
by allowing the most compact binary (or triple) to remain near the center
of mass of the halo and for other stars to return and scatter again and again.

We find that on average,
the number of HMXB candidates formed per stellar mass, $F_{\rm HMXB}$,
is a function of the number and orientation of close 3-body encounters.
Our results indicate that $F_{\rm HMXB}$ may be somewhat higher
in configurations that result in fewer ejections of stars, and if the interactions are coplanar.
While we are able to interpret this, as well as trends in the average
separation between stellar pairs, in terms of the initial kinematic setup
of the various scenarios simulated (see \S\ref{subsec:10-body} above), 
the value of $F_{\rm HMXB}$ does not vary by more than a factor $\approx 3$.
We interpret this lack of a significant variation in $F_{\rm HMXB}$,
for such a diverse set of initial conditions and ambient gas densities,
to mean that our values are not far from the one that results
from similar stellar encounters in nature.

\subsection{The effect of migration on the formation of HMXBs}\label{subsec:migration}

Another way in which a nascent stellar group could harden
is migration through a gaseous disk.
The migration could occur as the protostars form---\citet{Greif12} 
found significant accretion from the protostellar disk onto the most massive
protostar, and did not follow the evolution of the system beyond this stage.
(It could also occur to a lesser degree in a vestigial gas disk, after the stars are in place.)

We evaluate the possible role of disk migration on the separation of Pop~III
stars by considering a steady, geometrically thin disk with an $\alpha$ viscosity
(\citealt{SSdisk}; see also \citealt{Frank2002accretion}).
We adopt a disk with $\alpha=0.01$ and an accretion rate
$\dot{m}\sim 10^{-3}~\Msol\yr^{-1}$,
following \citet{Tan2004a} and \citet{Tan2004b},
who considered the structure of accretion disks around Pop~III
stars at high redshifts.

We estimate the migration timescale $\tau_{\rm mig}$ following \citet{Syer1995}.
We take a binary system with primary mass $M_{1}=120\Msol$
and secondary mass $M_{2}=11\Msol$, based on the mean values
found across our simulations.
For these masses and disk parameters, the secondary is able
to clear a gap around its orbital path
\citep[e.g.][]{Syer1995, Seager2010exoplanet, Lubow2010},
by satisfying both of the following two conditions:
\begin{equation}
\label{eq26}
\text{(1)} \,\ \frac{H}{R}\leq \Big(\frac{q}{\alpha}\Big)^{1/2},
\end{equation}
\begin{equation}
\label{eq21}
\text{(2)} \,\ \frac{H}{R}\leq \Big(\frac{q^{2}}{\alpha}\Big)^{1/5}.
\end{equation}
where $R$ is the distance of the secondary from the primary,
$H$ is the scale height of the disk at that location,
and $q=M_{2}/M_{1}\sim 0.1$ is the binary's mass ratio.
If condition (1) is violated, the gap will be closed by
the radial pressure gradient. Condition (2) relates the gap width and
the Roche radius of the secondary;
this ensures that the secondary acts to transfer orbital angular
momentum through the disk, rather than accreting mass via RLOF.
We find that inside $R\sim 15\AU$, for the disk properties stated above,
$H/R$ is smaller than the right-hand side of equation (\ref{eq26})
by a factor $\simgt 80$,
and smaller than the right-hand side of equation (\ref{eq21})
by a factor $\simgt 8$,
indicating that the secondary is easily able to open a gap.

The migration timescale of the secondary depends on the dimensionless parameter
\citep{Syer1995}
\begin{equation}
\label{eq27}
B\equiv\frac{4 \pi \Sigma_{0} R^{2}}{M_{2}},
\end{equation}
where $\Sigma_{0}$ is the local surface density of a steady-state
disk around the primary, in the absence of perturbations by the secondary.
For $B>1$, the gas in the disk is able to dynamically dominate
over the gravitational influence of the secondary, and the secondary
is pushed inward on the viscous diffusion timescale of the disk,
\begin{equation}
\label{eq20}
\tau_{{\rm mig},0}\sim \alpha^{-1}\Big(\frac{H}{R}\Big)^{-2}\Omega^{-1}\,.
\end{equation}
For $B<1$, \citet{Syer1995} found that the migration timescale
is longer,
\begin{equation}
\tau_{{\rm mig},1}\sim \frac{1}{B^{7/17}}\,\tau_{\text{mig,0}}\,.
\end{equation}

\begin{figure}
	\centering
	{\includegraphics[width=9.1cm]{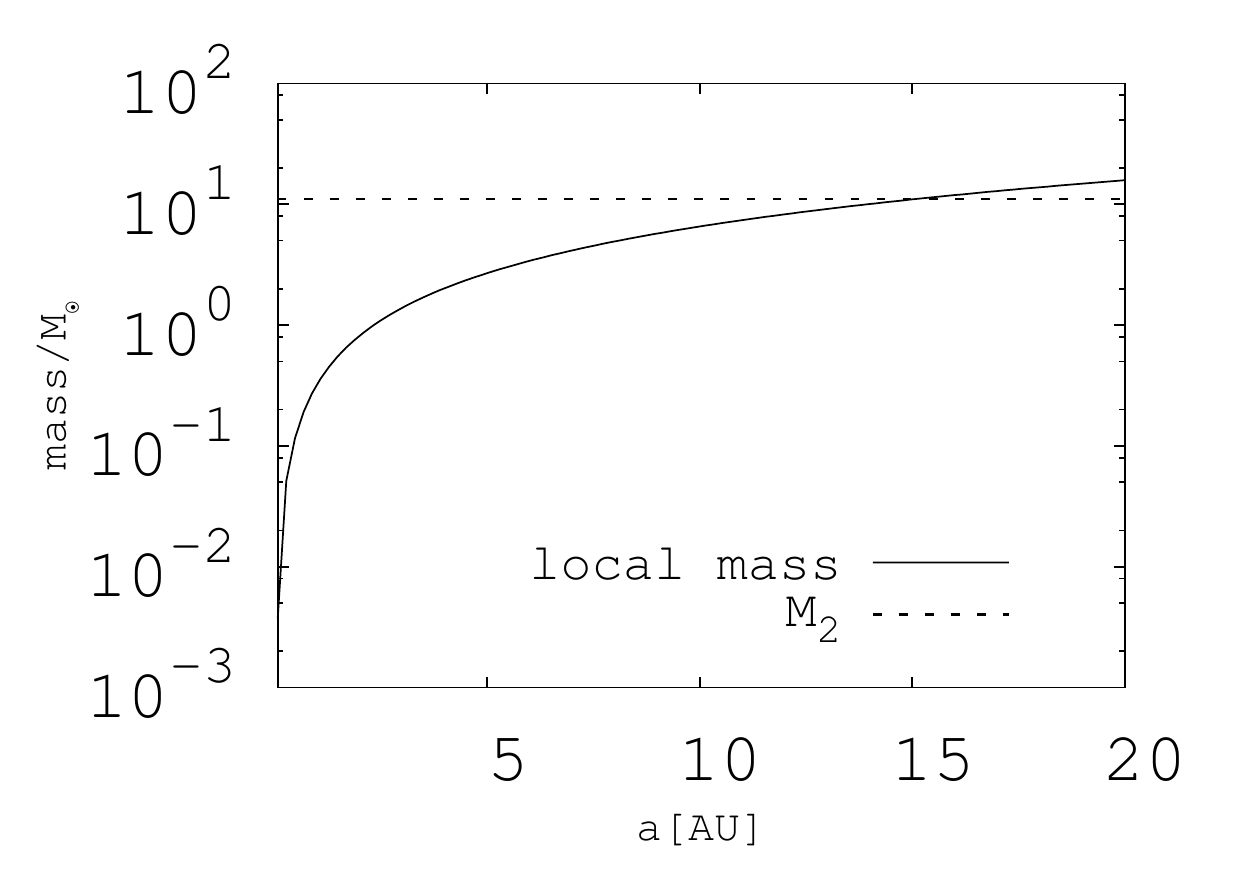}}
	\caption{A comparison between the mass of a typical
	secondary of a binary in our simulations ($M_{2}=11\Msol$,
	dashed line)
	and the local mass $4 \pi \Sigma_{0} R^{2}$ (solid line)
	in a protoplanetary disk near its orbit.
	 The migration timescale increases if the ratio
             $B=4 \pi\Sigma_{0} R^{2}/{M}_{2} <1$, which is the case for close orbits.}
	\label{fig:6}
\end{figure}

\begin{figure}
 	\centering
 	{\includegraphics[width=8.9cm]{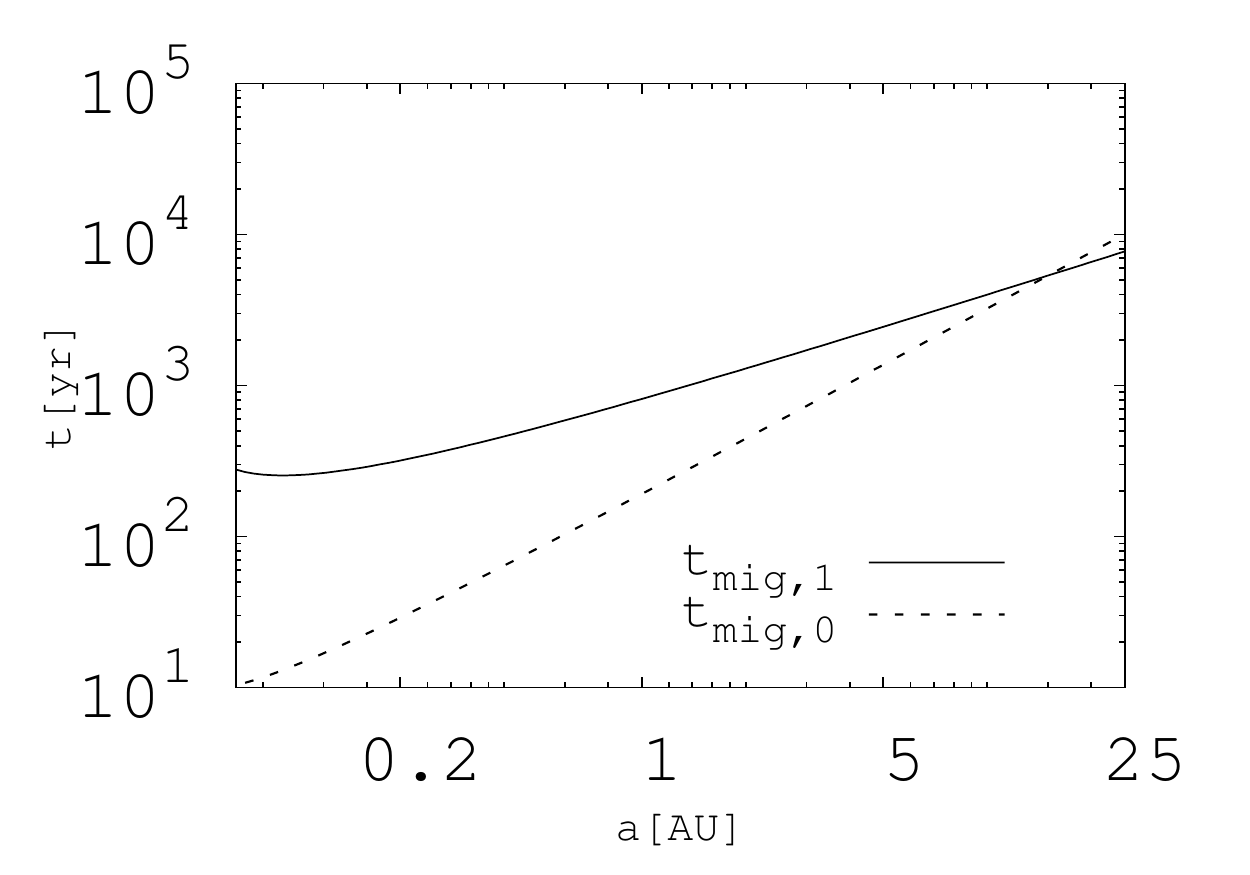}}
 	\caption{
	The migration timescales for a circumbinary
	protostellar disk, based on the typical binary properties of our
	simulations. 
	The unperturbed Type II migration timescale, $\tau_{{\rm mig},0}$ (dashed line)
	shows the timescale assuming that the secondary mass is much
	smaller than the local disk mass.
	A longer timescale $\tau_{{\rm mig},1}$ (solid line) is expected if
	the secondary mass is large compared to the disk mass (see previous Figure).}
 	\label{fig:9}
 \end{figure}

In Figure~\ref{fig:6}, we plot the local disk mass near the secondary,
$4 \pi \Sigma_{0} R^{2}$, alongside the typical secondary mass, $11\Msol$
(the numerator and denominator, respectively, for the ratio $B$).
For $R\simlt 15\AU$, $B<1$ and the migration timescale is expected to slow.
We plot the two migration timescales $\tau_{{\rm mig},0}$ and $\tau_{{\rm mig},1}$
in Figure~\ref{fig:9}.

Both of these timescales are shorter than both the typical lifetimes
of protostellar disks, as well as of the stars themselves.
This suggests that disk migration could, in principle, lead to initial
stellar separations smaller than what we have assumed, making
the formation of HMXBs via stellar scatterings more favorable.
On the other hand, radiative feedback from the stars could
blow away the disk before significant migration can occur.
We submit that the HMXB formation rates inferred from our simulations
be taken as a conservative estimate, with
possible additional contributions from channels
other than stellar scattering.

\subsection{X-ray output}
\label{subsec:X-rayspectra}
As discussed in \S\ref{sec:intro}, HMXBs are believed to be a major
source of X-rays in the early universe. 
Observations of nearby star-forming galaxies
suggest that their X-ray luminosities (which are dominated by HMXBs)
scale linearly with their star formation rate
\citep{Grimm2003,Gilfanov2004,Persic2004,xraysource2}.
\citet{Mineo2012} find 
many studies support a linear
proportionality between the X-ray luminosity of HMXBs and the star formation
rate (SFR) (\citealt{Grimm2003}, \citealt{Gilfanov2004},
\citealt{Persic2004} and \citealt{xraysource2}
). In the linear regime, the X-ray luminosity of
the local universe is given by \citet{Mineo2012},
\begin{equation}
\label{eq29}
L^{{\rm local}}_{2-10~{\rm keV}}=3\times 10^{39}\times \frac{\rm SFR}{\Msol \yr^{-1}},
\end{equation}
where SFR is the star formation rate.

How the ratio of X-ray luminosity to SFR evolves with redshift
is a key question in evaluating the properties of galaxies and young stellar populations,
and in linking the earliest galaxies (most of which should be actively forming stars,
based on their mass accretion rates) with their X-ray luminosities.
The question is as yet unresolved by current observations \citep{Dijkstra+12, BasuZych+13},
and is often treated as a free parameter in studies estimating the X-ray
production of the first galaxies \citep[e.g.][]{Furlanetto06, xrays3, Tanaka+15}.

In the following, based on the results of our simulations, 
we are going to quantitatively evaluate the relation between X-ray
luminosity and SFR, and compare it with equation (\ref{eq29}). We can write
the X-ray luminosity as
\begin{align}
\label{eq28}
L_{2-10~{\rm keV}}&=L_{\rm Edd}\times f_{\rm Edd}\times f_{2-10~{\rm keV}}\nonumber\\
& \qquad \times t_{{\rm acc}} \times f_{{\rm sur}}\times f_{{\rm esc}}\times F_{\rm HMXB} \times {\rm SFR}\,.
\end{align}

Below, we discuss each quantity in equation (\ref{eq28}).
\begin{enumerate}

\item $L_{\rm Edd}$, the Eddington luminosity, which scales with
the typical mass of the BH engine $M_{\rm BH}$ as
$1.3\times 10^{38} (M_{\rm BH}/\Msol)~\erg\s^{-1}$.

\item 
$f_{\rm Edd}$, the typical ratio of the total radiative power emitted by HMXBs
(the bolometric luminosity) to $L_{\rm Edd}$.
If the typical luminosity of a HMXB during an active phase is
Eddington, then $f_{\rm Edd}$ is effectively the mean duty cycle.
Other studies have adopted values ranging from $0.1$ to $0.5$
\citep{xraysource2, Belczynski2008, Salvaterra2012};
we take as our fiducial value $f_{\rm Edd}=0.1$.

\item

$f_{2-10~{\rm keV}}$, the fraction of the bolometric luminosity
that is emitted between $2$ and $10\keV$.
Observational estimates vary between $0.1$ and $0.8$
\citep[e.g.][]{Sipior2003, Migliari2006}.
Because the BH masses in Pop~III HMXBs are
expected to be higher than in present-day populations,
and the peak energies of accretion disks scale with the mass of the central engine as $M_{\rm BH}^{-1/4}$,
their characteristic spectra could be somewhat softer. 
We do not expect this to significantly affect our estimates here.

\item
$t_{{\rm acc}}$, the time that a massive binary spends as a HMXB,
with the less massive star donating mass to the more massive BH companion.
If the two stars form simultaneously and form a compact binary before the
more massive member dies to become a BH,
then this is simply $t_{2}-t_{1}$, the difference in the lifetimes of the stars.
We use this (somewhat arbitrary) estimate.
However, because the lifetimes of the less massive star
($t_{2}\simgt 10\Myr$ for stars with masses $\simlt 10\Msol$)
are comparable to the expected specific star formation rate in galaxies
at this redshift, we argue that any prescription that satisfies
$t_{{\rm acc}}\simlt t_{2}$ is a reasonable order-of-magnitude estimate.

\item $f_{{\rm sur}}$, the fraction of HMXB candidates identified
in our simulations that actually survive to become HMXBs.
This quantity accounts for possible disruptions of binaries,
due to (a) the merger of the stars during
main sequence and post-main sequence evolution \citep{Power2009};
(b) the more massive star getting kicked following a supernova explosion 
\citep[e.g.][]{Repetto2012, Janka2013};
and (c) subsequent disruptions by stellar scatterings that were not 
captured by our simulations.
(Our simulations follow the evolution of star groups until
the formation of a stable compact binary, but not until stellar death
millions of years later.)
Theoretical estimates typically yield $\sim 0.2-0.3$ \citep[e.g.][]{HMXB3, Artale2015}

\item $f_{{\rm esc}}$, the fraction of X-rays that escape the galaxy.
Unless the environment of the HMXBs are Compton-thick, which is
unlikely for the low-mass galaxies of interest, we expect $f_{{\rm esc}} \simlt 1$.

\item $F_{\rm HMXB}$, the number of HMXBs formed per stellar mass.
This is the main output of our simulations.
Whereas previous theoretical works had arrived at this value
by extrapolating the locally observed value with an assumed redshift evolution,
or with free parameters, here we provide an estimate based
on suites of $N$-body simulations whose initial conditions
are motivated by cosmological simulations of Pop~III star formation.
(Note that this quantity has units $\Msol^{-1}$; we use the capital letter
to distinguish it from the dimensionless fractions represented by $f$)

\end{enumerate}

Across all of our small-scale simulations---varying the number of stars in the group,
whether groups evolved in isolation or through several different orientations
of mutual collisions, and exploring two values for the ambient gas density
separated by two orders of magnitude---we find
$F_{{\rm HMXB}}\sim10^{-3}$, varying by less than a factor of $4$
between the lowest and the highest values (see Table \ref{tab:tab4}).

Finally, we can write $L_{2-10~{\rm keV}}$ as follows :  

\begin{align}
\label{eq30}
\frac{L_{2-10~{\rm keV}}}{{\rm SFR}}
   	=0.33&\times 
	\frac{M_{\rm BH}}{\Msol}\times \frac{f_{\rm Edd}}{0.1}\times \frac{f_{2-10~{\rm keV}}}{0.1}
	 \times \frac{t_{{\rm acc}}}{\rm Myr}\nonumber\\
	&
	\times \frac{f_{{\rm sur}}}{0.5}
	\times \frac{f_{{\rm esc}}}{0.5}\times \frac{F_{\rm HMXB}}{10^{-3}\Msol^{-1}}
	\nonumber\\
&\times \frac{10^{39}\erg\s^{-1}}{\Msol \yr^{-1}}\,.
\end{align}
Based on our choices of the factors, $f_{\rm Edd}=0.1$, $f_{2-10~{\rm
    keV}}=0.1$, $f_{{\rm sur}}=0.5$ and $f_{{\rm esc}}=0.5$, we can
estimate the normalized X-ray luminosities per SFR, $ L_{2-10 {\rm
    keV}}/{\rm SFR}$. 
    We report this quantity for each of our models in 
Table~\ref{tab:tab4}. It varies from a minimum of 37 to a maximum of 450 among
the studied scenarios.

These $L_{\rm X}$-to-SFR ratios are $\sim 40-150$ higher
  than what is observed in the local Universe.  This result is
  qualitatively consistent with the findings of \citet{BasuZych+13}
  and \citet{Kaaret2014}, who find an increase in the $L_{\rm
    X}$-to-SFR ratios toward $z\simgt 4$.
  Our high $L_{\rm X}$-to-SFR values stem from the large mass of the
  HMXBc primary, and the relatively low mass of the secondary.
  The former leads to a higher Eddington luminosity compared
  to typical stellar-mass BHs ($\sim 3\Msol$) in the local Universe,
  and the latter results in long stellar lifetimes, which in turn leads
  to longer $t_{\rm acc}$.
  Note that we used the stellar mass as a proxy for the BH mass
  for simplicity, due to the theoretical uncertainties in evaluating
  the mass loss due to winds and during the transition to a BH.
  Any significant mass loss (e.g. simulations by \citealt{ZhangW+08}
  suggest that the BH remnants of massive metal-free stars 
  end up with $\sim 40\%$ of the progenitor mass)
  would be directly translatable
  to the estimated X-ray luminosities reported in Table \ref{tab:tab4}.

\subsection{Implications for the thermal history of the IGM and the 21cm radiation}
\label{subsec:implicationtothermalhistory}

A higher $L_{\rm X}$-to-SFR ratio implies that IGM heating will
occur earlier than commonly thought.
The thermal history of the IGM can be probed in the $21\cm$ line, which
is observable in absorption (or in emission),
depending on whether the spin temperature of the IGM
is below (or above) the CMB temperature.

If the IGM heats early, as suggested by our estimates of the X-ray
emission of early galaxies, the 21~cm absorption line appears earlier,
and the ``dip'' as a function of redshift caused by adiabatic cooling
is not as deep and not as sharp as in the case of late heating as it
would otherwise (see Figure~2 in \citealt{Fialkov2014}).  Another
consequence of early, intense heating is that the temperature of the
IGM could become high enough, to suppress the formation of low-mass
galaxies \citep{Ripamonti+08} and the growth of their nuclear BHs
\citep{Tanaka2012}.

\begin{table*}
	\centering
	\setlength\extrarowheight{2pt}
	
	\begin{tabulary}{1\linewidth}{c | c c | c c c c c c}
		\hline
		& 5-body($n_{6})$ & 5-body($n_{4}$) & \textit{Sce.~1} & \textit{Sce.~2} & \textit{Sce.~3} & \textit{Sce.~4A} & \textit{Sce.~4B}\\
		\hline
		$M_{1}$[$\Msol$]($\tau_{\rm{life},1}[\Myr]$)	  &		110(2.5)					&	88(2.9) &	110(2.5)	& 		120(2.4) & 110(2.4) &  160(2.0) & 110(2.4)	\\
		$M_{2}$[$\Msol$] ($\tau_{\rm{life},2}[\rm{Myr}]$)	    &  	12(17)				&  11(18)	&		18(11)& 45(4.8)	& 28(7.1)& 63(3.7) & 63 (3.7) \\
		$t_{\rm{acc}}(=\tau_{\rm{life},2}-\tau_{\rm{life},1})$[Myr]   &  	14	&	15  & 8.6 	& 2.5 &	4.7 &	1.7 & 1.3  \\
		$F_{\rm HMXB}$ [$10^{-4}\Msol^{-1}$]	 &   4.6	&   4.6		&  15 &	11 & 9.0 & 4.2 & 8.4\\
		$L_{2-10 {\rm keV}}/{\rm SFR}$ $[10^{39}\rm{ erg s}^{-1}\Msol^{-1}\yr$] 	&	220 & 200 & 450 & 107 & 160 & 37 & 40 \\
		$L_{2-10 {\rm keV}}/L^{\rm local}_{2-10 {\rm keV}}$	&	75 & 67&  150 & 36 & 52 & 12 &  13\\
		\hline
		\hline
		$\frac{\eta_{\rm{GRB,Pop III}}}{\eta_{\rm{GRB,Pop I/II}}}$ &	2.2 & 2.2 & 7.0 & 5.4 & 4.3 & 2.0  & 4.0\\
		\hline
	\end{tabulary}
	
	\caption{Summary of the results of X-ray luminosity and GRB
          efficiencies. In the table, second and third columns
          correspond to 5-body calculations and the rest columns (from
          \textit{Sce.~1} to \textit{Sce.~4B}) are the results from
          10-body calculations with the number density of
          $10^{6}\rm{cm}^{-3}$.According to the ratio $L_{2-10~{\rm
              keV}}/L^{\rm local}_{2-10~{\rm keV}}$, where $L^{\rm
            local}_{2-10~{\rm keV}}=3\times10^{39}\erg\s^{-1}$, it is
          expected that the 2-10~keV X-ray luminosity is $\sim 10^{2}$
          times larger than the X-ray luminosity of the local
          universe. These larger values essentially result from the
            large mass of the star 1.  In the bottom row, we have
          listed the LGRB efficiency, derived from each set of
          simulations. The higher HMXB formation rates, lead to larger
          efficiencies than the estimated value for Pop~I/II stars,
          $\eta_{\rm{GRB,Pop I/II}}\simeq 4.2\times10^{-6}$ with
          beaming factor of $1/50$ (e.g. \citealt{Bromm2006}). }
	\label{tab:tab4}
\end{table*}

\subsection{Implications for Gamma Ray Bursts from Pop~III stars}
\label{subsec:implicationtoGRB}

As discussed in \S\ref{sec:intro},
the fraction of HMXBs at high redshifts has potential implications
for the expected rates of LGRBs from Pop~III stars.
According to the collapsar model \citep{MacFadyen1999},
for an exploding massive star to yield a
GRB, several conditions need to be satisfied, namely:
\begin{enumerate}
\item  The core of the star must collapse to a BH. This is realized
by most Pop~III stars, given their large masses.
\item The hydrogen envelope of the progenitor star must be
stripped, so that a relativistic jet can penetrate and exit the remaining envelope.
\item The BH should be surrounded by an accretion disk
of high angular momentum material. This is realized if the core of the
progenitor star has retained sufficient angular momentum during the
evolution.
\end{enumerate}
Binary systems 
more easily satisfy the last two conditions with respect to single stars
\citep[see e.g.][]{Cantiello2007}.
In fact, for single stars to end their lives as LGRBs, they need to be
born with large initial rotation (since they are less likely to be spun up by other stars),
and also avoid being slowed down by magnetic torques
\citep[e.g.][]{Spruit2002, Yoon2006, Perna2014}.

In contrast, binary stars can spin up the helium core of the progenitor star
via tidal coupling and spin-orbit locking.
Further, RLOF can strip the hydrogen envelope during a common-envelope
phase without reducing the rotation of the helium core \citep{Bromm2006}.
This is especially important for Pop~III stars,
whose heavier hydrogen envelopes would be more difficult to shed in isolation.
Therefore, compact binary systems, or HMXBs, constitute a
promising channel to produce LGRBs from Pop~III stars.
 
We can use our results for the formation rates of Pop~III HMXBs
to estimate the fraction of LGRBs from Pop~III stars\footnote{Note that in our study,
  since we are considering HMXB 'candidates', there is also the
  possibility of merger of the two stars during the common-envelope
  inspiral phase, when they both are stripped down to their Helium
  cores. This event could provide another avenue for the formation of
  GRBs \citep{Fryer2005}.  Alternatively, after both stars
  have undergone the SN explosion, if the system is still bound, the
  compact objects of the binary, upon merger as a result of
  gravitational energy loss, would be likely contributors to the
  population of Short Gamma-Ray Bursts
  (e.g. \citealt{Narayan1992}).}.
  \citet{Bromm2006} quantified the GRB
formation efficiency as
\begin{equation}
\eta_{\rm{GRB}}\simeq\eta_{\rm{BH}}\,\eta_{\rm{bin}}\,\eta_{\rm{close}}\,
\eta_{\rm{beaming}},
\end{equation}
where $\eta_{\rm{BH}}$ is the number of
BH-forming stars resulting from a given total stellar mass,
$\eta_{\rm{bin}}$ is the binary fraction and $\eta_{\rm{close}}$ is
the fraction of sufficiently close binaries to undergo RLOF.
For Pop~I/II stars they calculated $\eta_{\rm BH}\simeq 1/(700\Msol)$.
Combining this value with adopted values for the other parameters---$\eta_{\rm bin}\sim 0.5$,
$\eta_{\rm close}\sim 0.3$, and
$\eta_{\rm{beaming}}\simeq ({1}/{50})-({1}/{500})$---yields $\eta_{\rm GRB,PopI/II} \sim
4.2\times (10^{-6} - 10^{-7}) \Msol^{-1}$.

\citet{Bromm2006} noted that it was only to make educated guesses
for the Pop~III case, due to an absence of detailed
calculations for the fraction of close binaries.
Our work is a first attempt to fill the gap in our theoretical knowledge.
We can write
$F_{\rm HMXB}=\eta_{\rm{BH}}\,\eta_{\rm{bin}}\,\eta_{\rm{close}}$.
Adopting for comparison the same value of $\eta_{\rm{beaming}}\simeq ({1}/{50})-({1}/{500})$,
we then infer
$\eta_{\rm{GRB, Pop~III}}\simeq 4.8\times(10^{-6}\sim10^{-7})
\Msol^{-1}$ for the interacting 5-star case, and $2.2\times
(10^{-5}\sim10^{-6}) \Msol^{-1}$ for the (most favorable) 10-body
scenario.

Therefore, our results suggest that LGRB rates from Pop~III stars could be comparable to
or somewhat higher than the rates from Pop~I/II stars.

\subsection{Caveats}
\label{subsection:caveat}

Our suite of $N$-body simulations, spanning diverse sets of initial conditions
for Pop~III stars and their environments, point that HMXBs form
in higher fractions in the earliest galaxies than at low redshifts,
and make significant contributions to the thermal history of the IGM,
the $21\cm$ signature at $z\sim 20$, and to the rates of LGRBs.
The fact that the formation rates of HMXBs varied little between the simulations
suggest that our estimates for $F_{\rm HMXB}$ is reasonably robust.
However,  we here point out several uncertainties of our work that could
affect our conclusions.

One important factor is the IMF of the stars \citep[e.g.][]{Hirano+14}.
We adopted the IMF of \citet{stacy13}, which were based
on the masses of protostars roughly $5000\yr$ after the formation
of the protostellar seeds.

However, our key results are based on fragmentation of the protostellar
clouds on smaller scales, as found in the simulations by \citet{greif11}.
In those simulations, the most massive protostars had the highest accretion
rates, suggesting that the IMF slope may be steeper than what we assumed.
We also did not account for changes in the masses of stars as they evolved.
These are important considerations, as the masses of the stars play
a critical role in determining the time that a binary spends transferring mass
as a HMXB. 

 We made an effort to account for this limitation by using free parameters such as
 $f_{{\rm sur}}$, which accounts for the mass loss during the SN explosion.

Another factor that could be more carefully treated in future studies
is the spin of the star.  In a binary system, this impacts the
circularization of the orbit, followed by low eccentricity and the
synchronization of spin with orbital phase.\footnote{These effects had
  been often assumed to be due to the tidal interations with accreting
  gas, but recent studies suggest that the orbital semimajor axis and
  eccentricity can either increase or decrease depending on the binary
  properties at pericenter (\citealt{Sepinsky2007} and
  \citealt{Sepinsky2009}).}  Due to tidal dissipation and
circularization, this could additionally decrease the orbital distance
of a binary, boosting the formation of HMXBs and change the average
value of the eccentricity.

In addition to reducing the orbital
separation, the spin makes a difference in mass transfer rates.

Our simulations generally produce binaries with large eccentricities.
For highly eccentric orbits, mass transfer occurs
only near pericenter (\citealt{Lajoie2011} and
\citealt{Sepinsky2010}), 
and the rates depend
on whether the orbital angular speed of the star is
super-sychronous or sub-synchronous with rotational angular speed
\citep{Davis2013}.

We found that the viscous timescales in our eccentric binaries
were longer than the orbital timescales in the majority of cases, and used this
fact to conclude that their duty cycle should be the same as for
nearly circular binaries. However, we did find one exception,
in which a RLOF accretion event would be sufficiently short-lived to
be episodic. A more careful study focused on stellar rotation effects
may be necessary to more conclusively estimate the duty cycle
of eccentric Pop~III HMXBs.

Finally, our simulations are an attempt to model the stellar dynamics
as a gravitational $N$-body problem with perturbative forces due to a
fixed, smooth gaseous background.  More detailed simulations that
include detailed stellar feedback, as well as the dynamics and
thermodynamics of the gaseous environment, could lead to additional
revelations about the early evolution of Pop~III star groups.

\section{Summary}\label{sec:summary}

In this study, we used $N$-body simulations of the first stars
to explore the formation, evolution, disruption and energy output
of Pop~III HMXBs.
The code includes gravitational scattering of stars,
dynamical friction, and the gravitational potential of ambient gas.

The initial conditions for the simulations
(i.e. IMF, typical star separation in the host haloes, ambient
densities) are taken from two different sets of cosmological
simulations of Pop~III formation, namely by \citet{stacy13} 
('large-scale', i.e. a few thousands of AU),  and \citet{Greif12},
('small-scale', i.e. a few tens of AU).  These provide two
complementary sets in that they explore different physical scales for
star formation (for details, see \ref{sec:initialsetup}).  For each of
the two scenarios, we investigated star evolution in two backgound gas
densities, a high-density case ($10^6 {\rm cm}^{-3}$), and a
lower-density one ($10^4 {\rm cm}^{-3}$).

Based on the handful of protostars per halo that are found in the
works quoted above, we simulated systems with 5 stars and systems with
10 stars. We found:
\begin{enumerate}
\item \textit{5-body simulations}:
If stars form in quasi-Keplerian disk configurations with initial separations of hundreds of AU, 
HMXBs are highly unlikely to form.
In contrast, if stars form in compact groups separated by $\sim 10\AU$,
as is expected from turbulent fragmentation,
stellar scatterings lead to a significant HMXB formation rate.
In particular, we found that HMXBs form at a rate of a few per $10^{4}\Msol$
of stars formed, independent of the ambient gas density.
\item \textit{10-body  simulations}:
We simulated 10 stars on separations of $\sim 10\AU$,
and evolved them as isolated quasi-Keplerian disks,
or as two colliding groups with 5 stars each.
For the latter, we ran several different sets of collision geometries.

We found that the HMXB formation rate was a factor $\sim 1-3$ times 
higher than for the $5$-body simulations, mainly due to the fact that
the larger number of stars allowed for more hardening via stellar scattering.
\end{enumerate}

All of the small-scale simulations suggest an X-ray luminosity per
unit star formation that is a factor $\sim 10^{2}$ higher than what is
observed in the local Universe (under the assumption that other
variables such as the X-ray escape fraction from galaxies and the duty
cycle of HMXBs do not differ significantly).  These results are mostly
due to \textit{the large mass of the most massive star of the HMXBc compared
to that of the companion star, implying both a large $t_{\rm acc}$ as
well as a higher luminosity of the remnant BH}.  The fact that we found
little variation in this quantity across all of our simulations
suggests that this is a robust estimate.  Additional factors, such as
in-disk migration of nascent stars, could further increase the HMXB
formation efficiency.

A direct consequence is that X-rays can heat the IGM rapidly at Cosmic Dawn.
Signals of early heating can be probed via
the 21~cm line radiation: the absorption line signal is expected to show
a broader, shallower  minimum due to the shorter gas cooling time,  while the emission
line would be observed earlier because of the higher gas temperature,
at earlier times.
Several studies have modeled the 21~cm signature of the first HMXBs,
but relied on assumptions for their $L_{\rm X}/{\rm SFR}$ relation relative
to the empirical value found at lower redshifts. Our work provides a theoretically
driven estimate for this quantity.

In addition to the
implications for the thermal history of the IGM, these high formation rates
of HMXBs per stellar mass imply a higher GRB formation efficiency from
Pop~III stars in binaries. This predictions can be tested with a long
baseline of observational data from \textit{Swift}.

\vspace{0.5cm}

{\bf Acknowledgements} We are grateful to Zolt\'an Haiman and Mark Dijkstra for meaningful comments and suggestions which helped to improve this work.


\begin{thebibliography}{104}
	\expandafter\ifx\csname natexlab\endcsname\relax\def\natexlab#1{#1}\fi
	
	\bibitem[{{Abel}, {Bryan} \& {Norman}(2002){Abel}, {Bryan}, \&
		{Norman}}]{Abel+02}
	{Abel} T., {Bryan} G.~L., {Norman} M.~L., 2002, Science, 295, 93
	
	\bibitem[{{Artale}, {Tissera} \& {Pellizza}(2015){Artale}, {Tissera}, \&
		{Pellizza}}]{Artale2015}
	{Artale} M.~C., {Tissera} P.~B., {Pellizza} L.~J., 2015, \mnras, 448, 3071
	
	\bibitem[{{Ba{\~n}ados} {et~al}\mbox{.}(2014){Ba{\~n}ados}, {Venemans},
		{Morganson}, {Decarli}, {Walter}, {Chambers}, {Rix}, {Farina}, {Fan},
		{Jiang}, {McGreer}, {De Rosa}, {Simcoe}, {Wei{\ss}}, {Price}, {Morgan},
		{Burgett}, {Greiner}, {Kaiser}, {Kudritzki}, {Magnier}, {Metcalfe}, {Stubbs},
		{Sweeney}, {Tonry}, {Wainscoat}, \& {Waters}}]{Banados+14}
	{Ba{\~n}ados} E. {et~al.}, 2014, \aj, 148, 14
	
	\bibitem[{{Basu-Zych} {et~al}\mbox{.}(2013){Basu-Zych}, {Lehmer},
		{Hornschemeier}, {Bouwens}, {Fragos}, {Oesch}, {Belczynski}, {Brandt},
		{Kalogera}, {Luo}, {Miller}, {Mullaney}, {Tzanavaris}, {Xue}, \&
		{Zezas}}]{BasuZych+13}
	{Basu-Zych} A.~R. {et~al.}, 2013, \apj, 762, 45
	
	\bibitem[{{Belczynski} {et~al}\mbox{.}(2008){Belczynski}, {Kalogera}, {Rasio},
		{Taam}, {Zezas}, {Bulik}, {Maccarone}, \& {Ivanova}}]{Belczynski2008}
	{Belczynski} K., {Kalogera} V., {Rasio} F.~A., {Taam} R.~E., {Zezas} A.,
	{Bulik} T., {Maccarone} T.~J., {Ivanova} N., 2008, \apjs, 174, 223
	
	\bibitem[{{Binney} \& {Tremaine}(1987)}]{Galacticdynamics}
	{Binney} J., {Tremaine} S., 1987, {Galactic dynamics}
	
	\bibitem[{{Bowman}, {Rogers} \& {Hewitt}(2008){Bowman}, {Rogers}, \&
		{Hewitt}}]{Bowman+08}
	{Bowman} J.~D., {Rogers} A.~E.~E., {Hewitt} J.~N., 2008, \apj, 676, 1
	
	\bibitem[{{Bromm} \& {Loeb}(2006)}]{Bromm2006}
	{Bromm} V., {Loeb} A., 2006, \apj, 642, 382
	
	\bibitem[{{Bromm} \& {Yoshida}(2011)}]{BrommYoshida11}
	{Bromm} V., {Yoshida} N., 2011, \araa, 49, 373
	
	\bibitem[{{Burns} {et~al}\mbox{.}(2012){Burns}, {Lazio}, {Bale}, {Bowman},
		{Bradley}, {Carilli}, {Furlanetto}, {Harker}, {Loeb}, \&
		{Pritchard}}]{DARE12}
	{Burns} J.~O. {et~al.}, 2012, Advances in Space Research, 49, 433
	
	\bibitem[{{Cantiello} {et~al}\mbox{.}(2007){Cantiello}, {Yoon}, {Langer}, \&
		{Livio}}]{Cantiello2007}
	{Cantiello} M., {Yoon} S.-C., {Langer} N., {Livio} M., 2007, \aap, 465, L29
	
	\bibitem[{{Davis}, {Siess} \& {Deschamps}(2013){Davis}, {Siess}, \&
		{Deschamps}}]{Davis2013}
	{Davis} P.~J., {Siess} L., {Deschamps} R., 2013, \aap, 556, A4
	
	\bibitem[{{Dijkstra} {et~al}\mbox{.}(2012){Dijkstra}, {Gilfanov}, {Loeb}, \&
		{Sunyaev}}]{Dijkstra+12}
	{Dijkstra} M., {Gilfanov} M., {Loeb} A., {Sunyaev} R., 2012, \mnras, 421, 213
	
	\bibitem[{{Eggleton}(1983)}]{Rocheradius}
	{Eggleton} P.~P., 1983, \apj, 268, 368
	
	\bibitem[{{Erwin}(1969)}]{Fehlberg}
	{Erwin} F., 1969, NASA Technical Report, 315, 1
	
	\bibitem[{{Escala} {et~al}\mbox{.}(2004){Escala}, {Larson}, {Coppi}, \&
		{Mardones}}]{dyformula3}
	{Escala} A., {Larson} R.~B., {Coppi} P.~S., {Mardones} D., 2004, \apj, 607, 765
	
	\bibitem[{{Fan} {et~al}\mbox{.}(2001){Fan}, {Narayanan}, {Lupton}, {Strauss},
		{Knapp}, {Becker}, {White}, {Pentericci}, {Leggett}, {Haiman}, {Gunn},
		{Ivezi{\'c}}, {Schneider}, {Anderson}, {Brinkmann}, {Bahcall}, {Connolly},
		{Csabai}, {Doi}, {Fukugita}, {Geballe}, {Grebel}, {Harbeck}, {Hennessy},
		{Lamb}, {Miknaitis}, {Munn}, {Nichol}, {Okamura}, {Pier}, {Prada},
		{Richards}, {Szalay}, \& {York}}]{Fan+01}
	{Fan} X. {et~al.}, 2001, \aj, 122, 2833
	
	\bibitem[{{Fialkov} \& {Barkana}(2014)}]{Fialkov2014}
	{Fialkov} A., {Barkana} R., 2014, \mnras, 445, 213
	
	\bibitem[{{Fialkov}, {Barkana} \& {Visbal}(2014){Fialkov}, {Barkana}, \&
		{Visbal}}]{xrays3}
	{Fialkov} A., {Barkana} R., {Visbal} E., 2014, \nat, 506, 197
	
	\bibitem[{{Frank}, {King} \& {Raine}(2002){Frank}, {King}, \&
		{Raine}}]{Frank2002accretion}
	{Frank} J., {King} A., {Raine} D.~J., 2002, {Accretion Power in Astrophysics:
		Third Edition}
	
	\bibitem[{{Fryer} \& {Heger}(2005)}]{Fryer2005}
	{Fryer} C.~L., {Heger} A., 2005, \apj, 623, 302
	
	\bibitem[{{Furlanetto}(2006)}]{Furlanetto06}
	{Furlanetto} S.~R., 2006, \mnras, 371, 867
	
	\bibitem[{{Gilfanov}, {Grimm} \& {Sunyaev}(2004){Gilfanov}, {Grimm}, \&
		{Sunyaev}}]{Gilfanov2004}
	{Gilfanov} M., {Grimm} H.-J., {Sunyaev} R., 2004, \mnras, 347, L57
	
	\bibitem[{{Gnedin}(2000)}]{Gnedin00}
	{Gnedin} N.~Y., 2000, \apj, 542, 535
	
	\bibitem[{{Greif} {et~al}\mbox{.}(2012){Greif}, {Bromm}, {Clark}, {Glover},
		{Smith}, {Klessen}, {Yoshida}, \& {Springel}}]{Greif12}
	{Greif} T.~H., {Bromm} V., {Clark} P.~C., {Glover} S.~C.~O., {Smith} R.~J.,
	{Klessen} R.~S., {Yoshida} N., {Springel} V., 2012, \mnras, 424, 399
	
	\bibitem[{{Greif} {et~al}\mbox{.}(2011){Greif}, {Springel}, {White}, {Glover},
		{Clark}, {Smith}, {Klessen}, \& {Bromm}}]{greif11}
	{Greif} T.~H., {Springel} V., {White} S.~D.~M., {Glover} S.~C.~O., {Clark}
	P.~C., {Smith} R.~J., {Klessen} R.~S., {Bromm} V., 2011, \apj, 737, 75
	
	\bibitem[{{Grimm}, {Gilfanov} \& {Sunyaev}(2003){Grimm}, {Gilfanov}, \&
		{Sunyaev}}]{Grimm2003}
	{Grimm} H.-J., {Gilfanov} M., {Sunyaev} R., 2003, \mnras, 339, 793
	
	\bibitem[{{Haiman}(2013)}]{Haiman13}
	{Haiman} Z., 2013, in Astrophysics and Space Science Library, Vol. 396,
	Astrophysics and Space Science Library, {Wiklind} T., {Mobasher} B., {Bromm}
	V., eds., p. 293
	
	\bibitem[{{Haiman}, {Rees} \& {Loeb}(1996){Haiman}, {Rees}, \&
		{Loeb}}]{Haiman+96}
	{Haiman} Z., {Rees} M.~J., {Loeb} A., 1996, \apj, 467, 522
	
	\bibitem[{{Heger} {et~al}\mbox{.}(2003){Heger}, {Fryer}, {Woosley}, {Langer},
		\& {Hartmann}}]{Heger2003}
	{Heger} A., {Fryer} C.~L., {Woosley} S.~E., {Langer} N., {Hartmann} D.~H.,
	2003, \apj, 591, 288
	
	\bibitem[{{Hinshaw} {et~al}\mbox{.}(2013){Hinshaw}, {Larson}, {Komatsu},
		{Spergel}, {Bennett}, {Dunkley}, {Nolta}, {Halpern}, {Hill}, {Odegard},
		{Page}, {Smith}, {Weiland}, {Gold}, {Jarosik}, {Kogut}, {Limon}, {Meyer},
		{Tucker}, {Wollack}, \& {Wright}}]{Hinshaw+13}
	{Hinshaw} G. {et~al.}, 2013, \apjs, 208, 19
	
	\bibitem[{{Hirano} {et~al}\mbox{.}(2014){Hirano}, {Hosokawa}, {Yoshida},
		{Umeda}, {Omukai}, {Chiaki}, \& {Yorke}}]{Hirano+14}
	{Hirano} S., {Hosokawa} T., {Yoshida} N., {Umeda} H., {Omukai} K., {Chiaki} G.,
	{Yorke} H.~W., 2014, \apj, 781, 60
	
	\bibitem[{{Hopkins} \& {Quataert}(2010)}]{HopkinsQuataert10}
	{Hopkins} P.~F., {Quataert} E., 2010, \mnras, 407, 1529
	
	\bibitem[{{Inayoshi} \& {Tanaka}(2015)}]{Inayoshi2015}
	{Inayoshi} K., {Tanaka} T.~L., 2015, \mnras, 450, 4350
	
	\bibitem[{{Janka}(2013)}]{Janka2013}
	{Janka} H.-T., 2013, \mnras, 434, 1355
	
	\bibitem[{{Jeon} {et~al}\mbox{.}(2014){Jeon}, {Pawlik}, {Bromm}, \&
		{Milosavljevi{\'c}}}]{HMXB3}
	{Jeon} M., {Pawlik} A.~H., {Bromm} V., {Milosavljevi{\'c}} M., 2014, \mnras,
	440, 3778
	
	\bibitem[{{Kaaret}(2014)}]{Kaaret2014}
	{Kaaret} P., 2014, \mnras, 440, L26
	
	\bibitem[{{Kozai}(1962)}]{kozai}
	{Kozai} Y., 1962, \aj, 67, 591
	
	\bibitem[{{Kuhlen} \& {Madau}(2005)}]{Kuhlen2005}
	{Kuhlen} M., {Madau} P., 2005, \mnras, 363, 1069
	
	\bibitem[{{Lajoie} \& {Sills}(2011)}]{Lajoie2011}
	{Lajoie} C.-P., {Sills} A., 2011, \apj, 726, 67
	
	\bibitem[{{Latif} {et~al}\mbox{.}(2015){Latif}, {Bovino}, {Grassi},
		{Schleicher}, \& {Spaans}}]{Latif+15}
	{Latif} M.~A., {Bovino} S., {Grassi} T., {Schleicher} D.~R.~G., {Spaans} M.,
	2015, \mnras, 446, 3163
	
	\bibitem[{{Lubow} \& {Ida}(2010)}]{Lubow2010}
	{Lubow} S.~H., {Ida} S., 2010, {Planet Migration}, {Seager} S., ed., pp.
	347--371
	
	\bibitem[{{MacFadyen} \& {Woosley}(1999)}]{MacFadyen1999}
	{MacFadyen} A.~I., {Woosley} S.~E., 1999, \apj, 524, 262
	
	\bibitem[{{Machacek}, {Bryan} \& {Abel}(2003){Machacek}, {Bryan}, \&
		{Abel}}]{Machacek+03}
	{Machacek} M.~E., {Bryan} G.~L., {Abel} T., 2003, \mnras, 338, 273
	
	\bibitem[{{Madau}, {Haardt} \& {Dotti}(2014){Madau}, {Haardt}, \&
		{Dotti}}]{Madau+14}
	{Madau} P., {Haardt} F., {Dotti} M., 2014, \apjl, 784, L38
	
	\bibitem[{{Marigo}, {Chiosi} \& {Kudritzki}(2003){Marigo}, {Chiosi}, \&
		{Kudritzki}}]{Marigo}
	{Marigo} P., {Chiosi} C., {Kudritzki} R.-P., 2003, \aap, 399, 617
	
	\bibitem[{{McGreer} {et~al}\mbox{.}(2006){McGreer}, {Becker}, {Helfand}, \&
		{White}}]{McGreer+06}
	{McGreer} I.~D., {Becker} R.~H., {Helfand} D.~J., {White} R.~L., 2006, \apj,
	652, 157
	
	\bibitem[{{Migliari} \& {Fender}(2006)}]{Migliari2006}
	{Migliari} S., {Fender} R.~P., 2006, \mnras, 366, 79
	
	\bibitem[{{Mineo}, {Gilfanov} \& {Sunyaev}(2012){Mineo}, {Gilfanov}, \&
		{Sunyaev}}]{Mineo2012}
	{Mineo} S., {Gilfanov} M., {Sunyaev} R., 2012, \mnras, 419, 2095
	
	\bibitem[{{Mirabel} {et~al}\mbox{.}(2011){Mirabel}, {Dijkstra}, {Laurent},
		{Loeb}, \& {Pritchard}}]{xraysource2}
	{Mirabel} I.~F., {Dijkstra} M., {Laurent} P., {Loeb} A., {Pritchard} J.~R.,
	2011, \aap, 528, A149
	
	\bibitem[{{Mortlock} {et~al}\mbox{.}(2011){Mortlock}, {Warren}, {Venemans},
		{Patel}, {Hewett}, {McMahon}, {Simpson}, {Theuns}, {Gonz{\'a}les-Solares},
		{Adamson}, {Dye}, {Hambly}, {Hirst}, {Irwin}, {Kuiper}, {Lawrence}, \&
		{R{\"o}ttgering}}]{Mortlock+11}
	{Mortlock} D.~J. {et~al.}, 2011, \nat, 474, 616
	
	\bibitem[{{Naoz} \& {Barkana}(2007)}]{NaozBarkana07}
	{Naoz} S., {Barkana} R., 2007, \mnras, 377, 667
	
	\bibitem[{{Narayan}, {Paczynski} \& {Piran}(1992){Narayan}, {Paczynski}, \&
		{Piran}}]{Narayan1992}
	{Narayan} R., {Paczynski} B., {Piran} T., 1992, \apjl, 395, L83
	
	\bibitem[{{Oh}(2001)}]{Peng2001}
	{Oh} S.~P., 2001, \apj, 553, 499
	
	\bibitem[{{Ostriker}(1999)}]{dyformula2}
	{Ostriker} E.~C., 1999, \apj, 513, 252
	
	\bibitem[{{Pacucci} {et~al}\mbox{.}(2015){Pacucci}, {Ferrara}, {Volonteri}, \&
		{Dubus}}]{Pacucci+15}
	{Pacucci} F., {Ferrara} A., {Volonteri} M., {Dubus} G., 2015, ArXiv e-prints
	
	\bibitem[{{Perna} {et~al}\mbox{.}(2014){Perna}, {Duffell}, {Cantiello}, \&
		{MacFadyen}}]{Perna2014}
	{Perna} R., {Duffell} P., {Cantiello} M., {MacFadyen} A.~I., 2014, \apj, 781,
	119
	
	\bibitem[{{Persic} {et~al}\mbox{.}(2004){Persic}, {Rephaeli}, {Braito},
		{Cappi}, {Della Ceca}, {Franceschini}, \& {Gruber}}]{Persic2004}
	{Persic} M., {Rephaeli} Y., {Braito} V., {Cappi} M., {Della Ceca} R.,
	{Franceschini} A., {Gruber} D.~E., 2004, \aap, 419, 849
	
	\bibitem[{{Planck Collaboration} {et~al}\mbox{.}(2015){Planck Collaboration},
		{Ade}, {Aghanim}, {Arnaud}, {Ashdown}, {Aumont}, {Baccigalupi}, {Banday},
		{Barreiro}, {Bartlett}, \& et~al.}]{Planck15}
	{Planck Collaboration} {et~al.}, 2015, ArXiv e-prints
	
	\bibitem[{{Power} {et~al}\mbox{.}(2009){Power}, {Wynn}, {Combet}, \&
		{Wilkinson}}]{Power2009}
	{Power} C., {Wynn} G.~A., {Combet} C., {Wilkinson} M.~I., 2009, \mnras, 395,
	1146
	
	\bibitem[{{Pritchard} \& {Furlanetto}(2007)}]{Pritchard2007}
	{Pritchard} J.~R., {Furlanetto} S.~R., 2007, \mnras, 376, 1680
	
	\bibitem[{{Pritchard} \& {Loeb}(2008)}]{PritchardJ2008}
	{Pritchard} J.~R., {Loeb} A., 2008, \prd, 78, 103511
	
	\bibitem[{{Reg{\"o}s}, {Bailey} \& {Mardling}(2005){Reg{\"o}s}, {Bailey}, \&
		{Mardling}}]{masstransfer}
	{Reg{\"o}s} E., {Bailey} V.~C., {Mardling} R., 2005, \mnras, 358, 544
	
	\bibitem[{{Repetto}, {Davies} \& {Sigurdsson}(2012){Repetto}, {Davies}, \&
		{Sigurdsson}}]{Repetto2012}
	{Repetto} S., {Davies} M.~B., {Sigurdsson} S., 2012, \mnras, 425, 2799
	
	\bibitem[{{Ricotti} \& {Ostriker}(2004)}]{Ricotti2004}
	{Ricotti} M., {Ostriker} J.~P., 2004, \mnras, 352, 547
	
	\bibitem[{{Ripamonti}, {Mapelli} \& {Zaroubi}(2008){Ripamonti}, {Mapelli}, \&
		{Zaroubi}}]{Ripamonti+08}
	{Ripamonti} E., {Mapelli} M., {Zaroubi} S., 2008, \mnras, 387, 158
	
	\bibitem[{{Salvaterra} {et~al}\mbox{.}(2012{\natexlab{a}}){Salvaterra},
		{Haardt}, {Volonteri}, \& {Moretti}}]{Salvaterra+12}
	{Salvaterra} R., {Haardt} F., {Volonteri} M., {Moretti} A., 2012{\natexlab{a}},
	\aap, 545, L6
	
	\bibitem[{{Salvaterra} {et~al}\mbox{.}(2012{\natexlab{b}}){Salvaterra},
		{Haardt}, {Volonteri}, \& {Moretti}}]{Salvaterra2012}
	{Salvaterra} R., {Haardt} F., {Volonteri} M., {Moretti} A., 2012{\natexlab{b}},
	\aap, 545, L6
	
	\bibitem[{{Schaerer}(2002)}]{Schaerer}
	{Schaerer} D., 2002, in Astrophysics and Space Science Library, Vol. 274, New
	Quests in Stellar Astrophysics: the Link Between Stars and Cosmology,
	{Ch{\'a}vez} M., {Bressan} A., {Buzzoni} A., {Mayya} D., eds., pp. 185--188
	
	\bibitem[{{Seager}(2010)}]{Seager2010exoplanet}
	{Seager} S., 2010, {Exoplanets}
	
	\bibitem[{{Sepinsky}, {Willems} \& {Kalogera}(2007){Sepinsky}, {Willems}, \&
		{Kalogera}}]{Sepinsky2007}
	{Sepinsky} J.~F., {Willems} B., {Kalogera} V., 2007, \apj, 660, 1624
	
	\bibitem[{{Sepinsky} {et~al}\mbox{.}(2009){Sepinsky}, {Willems}, {Kalogera}, \&
		{Rasio}}]{Sepinsky2009}
	{Sepinsky} J.~F., {Willems} B., {Kalogera} V., {Rasio} F.~A., 2009, \apj, 702,
	1387
	
	\bibitem[{{Sepinsky} {et~al}\mbox{.}(2010){Sepinsky}, {Willems}, {Kalogera}, \&
		{Rasio}}]{Sepinsky2010}
	{Sepinsky} J.~F., {Willems} B., {Kalogera} V., {Rasio} F.~A., 2010, \apj, 724,
	546
	
	\bibitem[{{Shakura} \& {Sunyaev}(1973)}]{SSdisk}
	{Shakura} N.~I., {Sunyaev} R.~A., 1973, \aap, 24, 337
	
	\bibitem[{{Shankar}, {Weinberg} \& {Miralda-Escud{\'e}}(2009){Shankar},
		{Weinberg}, \& {Miralda-Escud{\'e}}}]{Shankar+09}
	{Shankar} F., {Weinberg} D.~H., {Miralda-Escud{\'e}} J., 2009, \apj, 690, 20
	
	\bibitem[{{Shen} {et~al}\mbox{.}(2007){Shen}, {Strauss}, {Oguri}, {Hennawi},
		{Fan}, {Richards}, {Hall}, {Gunn}, {Schneider}, {Szalay}, {Thakar}, {Vanden
			Berk}, {Anderson}, {Bahcall}, {Connolly}, \& {Knapp}}]{Shen+07}
	{Shen} Y. {et~al.}, 2007, \aj, 133, 2222
	
	\bibitem[{{Sijacki} {et~al}\mbox{.}(2007){Sijacki}, {Springel}, {Di Matteo}, \&
		{Hernquist}}]{Sijacki+07}
	{Sijacki} D., {Springel} V., {Di Matteo} T., {Hernquist} L., 2007, \mnras, 380,
	877
	
	\bibitem[{{Sipior}, {Eracleous} \& {Sigurdsson}(2003){Sipior}, {Eracleous}, \&
		{Sigurdsson}}]{Sipior2003}
	{Sipior} M.~S., {Eracleous} M., {Sigurdsson} S., 2003, ArXiv Astrophysics
	e-prints
	
	\bibitem[{{Springel}, {Di Matteo} \& {Hernquist}(2005){Springel}, {Di Matteo},
		\& {Hernquist}}]{Springel+05}
	{Springel} V., {Di Matteo} T., {Hernquist} L., 2005, \mnras, 361, 776
	
	\bibitem[{{Spruit}(2002)}]{Spruit2002}
	{Spruit} H.~C., 2002, \aap, 381, 923
	
	\bibitem[{{Stacy} \& {Bromm}(2013)}]{stacy13}
	{Stacy} A., {Bromm} V., 2013, \mnras, 433, 1094
	
	\bibitem[{{Stacy}, {Greif} \& {Bromm}(2010){Stacy}, {Greif}, \&
		{Bromm}}]{Stacy+10}
	{Stacy} A., {Greif} T.~H., {Bromm} V., 2010, \mnras, 403, 45
	
	\bibitem[{{Stinson} {et~al}\mbox{.}(2006){Stinson}, {Seth}, {Katz}, {Wadsley},
		{Governato}, \& {Quinn}}]{Stinson+06}
	{Stinson} G., {Seth} A., {Katz} N., {Wadsley} J., {Governato} F., {Quinn} T.,
	2006, \mnras, 373, 1074
	
	\bibitem[{{Syer} \& {Clarke}(1995)}]{Syer1995}
	{Syer} D., {Clarke} C.~J., 1995, \mnras, 277, 758
	
	\bibitem[{{Tan} \& {Blackman}(2004)}]{Tan2004b}
	{Tan} J.~C., {Blackman} E.~G., 2004, \apj, 603, 401
	
	\bibitem[{{Tan} \& {McKee}(2004)}]{Tan2004a}
	{Tan} J.~C., {McKee} C.~F., 2004, \apj, 603, 383
	
	\bibitem[{{Tanaka} \& {Haiman}(2009)}]{dyformula1}
	{Tanaka} T., {Haiman} Z., 2009, \apj, 696, 1798
	
	\bibitem[{{Tanaka}, {O'Leary} \& {Perna}(2015){Tanaka}, {O'Leary}, \&
		{Perna}}]{Tanaka+15}
	{Tanaka} T., {O'Leary} R., {Perna} R., 2015, \mnras submitted
	
	\bibitem[{{Tanaka}, {Perna} \& {Haiman}(2012){Tanaka}, {Perna}, \&
		{Haiman}}]{Tanaka2012}
	{Tanaka} T., {Perna} R., {Haiman} Z., 2012, \mnras, 425, 2974
	
	\bibitem[{{Tanaka}(2014)}]{Tanaka14}
	{Tanaka} T.~L., 2014, Classical and Quantum Gravity, 31, 244005
	
	\bibitem[{{Tanaka} \& {Li}(2014)}]{TanakaLi14}
	{Tanaka} T.~L., {Li} M., 2014, \mnras, 439, 1092
	
	\bibitem[{{Toma}, {Sakamoto} \& {M{\'e}sz{\'a}ros}(2011){Toma}, {Sakamoto}, \&
		{M{\'e}sz{\'a}ros}}]{Toma2011}
	{Toma} K., {Sakamoto} T., {M{\'e}sz{\'a}ros} P., 2011, \apj, 731, 127
	
	\bibitem[{{Treister} {et~al}\mbox{.}(2012){Treister}, {Schawinski}, {Urry}, \&
		{Simmons}}]{Treister+12}
	{Treister} E., {Schawinski} K., {Urry} C.~M., {Simmons} B.~D., 2012, \apjl,
	758, L39
	
	\bibitem[{{Turk}, {Abel} \& {O'Shea}(2009){Turk}, {Abel}, \&
		{O'Shea}}]{Turk+09}
	{Turk} M.~J., {Abel} T., {O'Shea} B., 2009, Science, 325, 601
	
	\bibitem[{{van Haarlem} {et~al}\mbox{.}(2013){van Haarlem}, {Wise}, {Gunst},
		{Heald}, {McKean}, {Hessels}, {de Bruyn}, {Nijboer}, {Swinbank}, {Fallows},
		{Brentjens}, {Nelles}, {Beck}, {Falcke}, {Fender}, {H{\"o}randel},
		{Koopmans}, {Mann}, {Miley}, {R{\"o}ttgering}, {Stappers}, {Wijers},
		{Zaroubi}, {van den Akker}, {Alexov}, {Anderson}, {Anderson}, {van Ardenne},
		{Arts}, {Asgekar}, {Avruch}, {Batejat}, {B{\"a}hren}, {Bell}, {Bell}, {van
			Bemmel}, {Bennema}, {Bentum}, {Bernardi}, {Best}, {B{\^i}rzan}, {Bonafede},
		{Boonstra}, {Braun}, {Bregman}, {Breitling}, {van de Brink}, {Broderick},
		{Broekema}, {Brouw}, {Br{\"u}ggen}, {Butcher}, {van Cappellen}, {Ciardi},
		{Coenen}, {Conway}, {Coolen}, {Corstanje}, {Damstra}, {Davies}, {Deller},
		{Dettmar}, {van Diepen}, {Dijkstra}, {Donker}, {Doorduin}, {Dromer}, {Drost},
		{van Duin}, {Eisl{\"o}ffel}, {van Enst}, {Ferrari}, {Frieswijk}, {Gankema},
		{Garrett}, {de Gasperin}, {Gerbers}, {de Geus}, {Grie{\ss}meier}, {Grit},
		{Gruppen}, {Hamaker}, {Hassall}, {Hoeft}, {Holties}, {Horneffer}, {van der
			Horst}, {van Houwelingen}, {Huijgen}, {Iacobelli}, {Intema}, {Jackson},
		{Jelic}, {de Jong}, {Juette}, {Kant}, {Karastergiou}, {Koers}, {Kollen},
		{Kondratiev}, {Kooistra}, {Koopman}, {Koster}, {Kuniyoshi}, {Kramer},
		{Kuper}, {Lambropoulos}, {Law}, {van Leeuwen}, {Lemaitre}, {Loose}, {Maat},
		{Macario}, {Markoff}, {Masters}, {McFadden}, {McKay-Bukowski}, {Meijering},
		{Meulman}, {Mevius}, {Middelberg}, {Millenaar}, {Miller-Jones}, {Mohan},
		{Mol}, {Morawietz}, {Morganti}, {Mulcahy}, {Mulder}, {Munk}, {Nieuwenhuis},
		{van Nieuwpoort}, {Noordam}, {Norden}, {Noutsos}, {Offringa}, {Olofsson},
		{Omar}, {Orr{\'u}}, {Overeem}, {Paas}, {Pandey-Pommier}, {Pandey}, {Pizzo},
		{Polatidis}, {Rafferty}, {Rawlings}, {Reich}, {de Reijer}, {Reitsma},
		{Renting}, {Riemers}, {Rol}, {Romein}, {Roosjen}, {Ruiter}, {Scaife}, {van
			der Schaaf}, {Scheers}, {Schellart}, {Schoenmakers}, {Schoonderbeek},
		{Serylak}, {Shulevski}, {Sluman}, {Smirnov}, {Sobey}, {Spreeuw}, {Steinmetz},
		{Sterks}, {Stiepel}, {Stuurwold}, {Tagger}, {Tang}, {Tasse}, {Thomas},
		{Thoudam}, {Toribio}, {van der Tol}, {Usov}, {van Veelen}, {van der Veen},
		{ter Veen}, {Verbiest}, {Vermeulen}, {Vermaas}, {Vocks}, {Vogt}, {de Vos},
		{van der Wal}, {van Weeren}, {Weggemans}, {Weltevrede}, {White}, {Wijnholds},
		{Wilhelmsson}, {Wucknitz}, {Yatawatta}, {Zarka}, {Zensus}, \& {van
			Zwieten}}]{LOFAR13}
	{van Haarlem} M.~P. {et~al.}, 2013, \aap, 556, A2
	
	\bibitem[{{Venemans} {et~al}\mbox{.}(2013){Venemans}, {Findlay}, {Sutherland},
		{De Rosa}, {McMahon}, {Simcoe}, {Gonz{\'a}lez-Solares}, {Kuijken}, \&
		{Lewis}}]{Venemans+13}
	{Venemans} B.~P. {et~al.}, 2013, \apj, 779, 24
	
	\bibitem[{{Venkatesan}, {Giroux} \& {Shull}(2001){Venkatesan}, {Giroux}, \&
		{Shull}}]{Venkatesan+01}
	{Venkatesan} A., {Giroux} M.~L., {Shull} J.~M., 2001, \apj, 563, 1
	
	\bibitem[{{Venkatesan}, {Tumlinson} \& {Shull}(2003){Venkatesan}, {Tumlinson},
		\& {Shull}}]{Venkatesan+03}
	{Venkatesan} A., {Tumlinson} J., {Shull} J.~M., 2003, \apj, 584, 621
	
	\bibitem[{{Volonteri}(2010)}]{Volonteri10}
	{Volonteri} M., 2010, \aapr, 18, 279
	
	\bibitem[{{Voytek} {et~al}\mbox{.}(2014){Voytek}, {Natarajan}, {J{\'a}uregui
			Garc{\'{\i}}a}, {Peterson}, \& {L{\'o}pez-Cruz}}]{SCIHI14}
	{Voytek} T.~C., {Natarajan} A., {J{\'a}uregui Garc{\'{\i}}a} J.~M., {Peterson}
	J.~B., {L{\'o}pez-Cruz} O., 2014, \apjl, 782, L9
	
	\bibitem[{{Willott} {et~al}\mbox{.}(2007){Willott}, {Delorme}, {Omont},
		{Bergeron}, {Delfosse}, {Forveille}, {Albert}, {Reyl{\'e}}, {Hill},
		{Gully-Santiago}, {Vinten}, {Crampton}, {Hutchings}, {Schade}, {Simard},
		{Sawicki}, {Beelen}, \& {Cox}}]{Willott+07}
	{Willott} C.~J. {et~al.}, 2007, \aj, 134, 2435
	
	\bibitem[{{Willott} {et~al}\mbox{.}(2009){Willott}, {Delorme}, {Reyl{\'e}},
		{Albert}, {Bergeron}, {Crampton}, {Delfosse}, {Forveille}, {Hutchings},
		{McLure}, {Omont}, \& {Schade}}]{Willott+09}
	{Willott} C.~J. {et~al.}, 2009, \aj, 137, 3541
	
	\bibitem[{{Yoon}, {Langer} \& {Norman}(2006){Yoon}, {Langer}, \&
		{Norman}}]{Yoon2006}
	{Yoon} S.-C., {Langer} N., {Norman} C., 2006, \aap, 460, 199
	
	\bibitem[{{Zhang}, {Woosley} \& {Heger}(2008){Zhang}, {Woosley}, \&
		{Heger}}]{ZhangW+08}
	{Zhang} W., {Woosley} S.~E., {Heger} A., 2008, \apj, 679, 639
	
\end{thebibliography}
\end{document}